\global\def\draftcontrol{0}

   \def\versionno{ ktthermo -- draft   }

\catcode`\@=11

\expandafter\ifx\csname draftcontrol\endcsname\relax\global\def\draftcontrol{0}
\fi

{\count255=\time\divide\count255 by 60
\xdef\hourmin{\number\count255}
\multiply\count255 by-60\advance\count255 by\time
\xdef\hourmin{\hourmin:\ifnum\count255<10 0\fi\the\count255}}
\def\draftdate{\number\month/\number\day/\number\year\ \ \ \hourmin }

\newcommand\makepapertitle{\par
  \begingroup
    \renewcommand\thefootnote{\@fnsymbol\c@footnote}%
    \def\@makefnmark{\rlap{\@textsuperscript{\normalfont\@thefnmark}}}%
    \long\def\@makefntext##1{\parindent 1em\noindent
            \hb@xt@1.8em{%
                \hss\@textsuperscript{\normalfont\@thefnmark}}##1}%
     \newpage
     \global\@topnum\z@   
     \@makepapertitle
     \thispagestyle{empty}\@thanks
  \endgroup
  \setcounter{footnote}{0}%
  \global\let\thanks\relax
  \global\let\makepapertitle\relax
  \global\let\@makepapertitle\relax
  \global\let\@thanks\@empty
  \global\let\@author\@empty
  \global\let\@date\@empty
  \global\let\@title\@empty
  \global\let\title\relax
  \global\let\author\relax
  \global\let\date\relax
  \global\let\and\relax
  \def\version{\let\version\@version\@gobble}
}
\def\@makepapertitle{%
  \newpage
   \ifnum\draftcontrol=1 {}
   \version\versionno
   \vskip 3em%
   \else
   \hfill\hbox to 3cm {\parbox{5cm}{\@pubnum}\hss}%
   \vskip 3em%
   \fi
   \begin{center}%
   \let \footnote \thanks
     {\LARGE {\@title}}%
     \vskip 1.5em%
     {\normalsize
       \lineskip .5em%
       \begin{tabular}[t]{c}%
         \@author
       \end{tabular}\par}%
     \vskip 1.5em%
     {\@bstract}%
     \end{center}%
     \vskip 1.5em
     \@date%
   \par
}

\gdef\@pubnum{}
\def\pubnum#1{%
  \gdef\@pubnum{#1}}

\gdef\@bstract{}
\def\Abstract#1{%
  \gdef\@bstract{%
   \parbox{\textwidth-0pc}{%
   \centerline{\bf Abstract}\penalty1000%
\kern.2cm%
\noindent
\renewcommand\baselinestretch{1.0}%
{#1}}}
}

\def\ps@paper{\let\@mkboth\@gobbletwo%
     \ifnum\draftcontrol=1
    \def\@oddfoot{\hbox to \textwidth{\tiny \versionno \hfil\tiny\draftdate}%
    \hskip -\textwidth \hbox to \textwidth{\hfil\rm\thepage\hfil}}%
     \else\def\@oddfoot{\hbox to \textwidth{\hfil\rm\thepage\hfil}}
     \fi
     \let\@evenfoot\@oddfoot
}

\def\body{\clearpage
          \pagestyle{paper}
    }

\def\@version#1{\ifnum\draftcontrol=1
\typeout{}\typeout{#1}\typeout{}
\vskip3mm\centerline{\hbox{\fbox{\normalsize{\tt DRAFT -- #1 -- }
                   {\draftdate}}}}\vskip3mm
\fi}
\let\version\@version
\long\def\eqlabel#1{\ifnum\draftcontrol=1
                    \tag@false  
                    \tag*{(\theequation) \hbox to -0.2cm{\hspace{0cm}\small{#1}\hss}}
                    \refstepcounter{equation}
                    \edef\@currentlabel{\theequation}
                    \ltx@label{#1}          
                    \else
                    \label{#1}
                    \fi
                    }
\let\st@bibitem\@bibitem
\let\st@lbibitem\@lbibitem
\ifnum\draftcontrol=1
  \def\@bibitem#1{%
    \st@bibitem{#1}\a@@label{#1}\ignorespaces}
  \def\@lbibitem[#1]#2{%
    \st@lbibitem[#1]{#2}\a@@label{#2}\ignorespaces}
  \def\a@@label#1{%
    \gdef\a@lab{\smash{\normalfont\small#1}}
    \ifvmode
      \if@inlabel
        \global\setbox\@labels\hbox{%
          \llap{\a@lab\let\a@lab\relax
                \kern\@totalleftmargin\kern\marginparsep}%
          \box\@labels}%
      \fi
    \fi}
\fi

\documentclass[12pt,letterpaper]{article}

\usepackage{amsmath,amssymb,array,calc,epsfig}
\usepackage{psfrag,verbatim,bm}

\ifnum\draftcontrol=1
\fi

\renewcommand\baselinestretch{1.25}
\renewcommand\section{\@startsection {section}{1}{\z@}%
                                   {-3.5ex \@plus -1ex \@minus -.2ex}%
                                   {2.3ex \@plus.2ex}%
                                   {\normalfont\large\bfseries}}
\renewcommand\subsection{\@startsection{subsection}{2}{\z@}%
                                   {-3.25ex\@plus -1ex \@minus -.2ex}%
                                   {1.5ex \@plus .2ex}%
                                   {\normalfont\normalsize\bfseries}}
\renewcommand\subsubsection{\@startsection{subsubsection}{3}{\z@}%
                                   {-3.25ex\@plus -1ex \@minus -.2ex}%
                                   {1.5ex \@plus .2ex}%
                                   {\normalfont\normalsize\it}}
\renewcommand\paragraph{\@startsection{paragraph}{4}{\z@}%
                                   {-3.25ex\@plus -1ex \@minus -.2ex}%
                                   {1.5ex \@plus .2ex}%
                                   {\normalfont\normalsize\bf}}

\numberwithin{equation}{section}

\def\ie{{\it i.e.}}

\def\revise#1       {\raisebox{-0em}{\rule{3pt}{1em}}%
                     \marginpar{\raisebox{.5em}{\vrule width3pt\
                     \vrule width0pt height 0pt depth0.5em
                     \hbox to 0cm{\hspace{0cm}{%
                     \parbox[t]{4em}{\raggedright\footnotesize{#1}}}\hss}}}}

\newcommand\nxt[1]  {\\\fnxt#1}

\def\cale         {{\cal E}}
\def\calf         {{\cal F}}

\def\calj         {{\cal J}}

\def\calo         {{\cal O}}
\def\calp         {{\cal P}}

\def\del          {\partial}

\def\be{\begin{equation}}
\def\ee{\end{equation}}
\def\bea{\begin{eqnarray}}
\def\eea{\end{eqnarray}}

\def\sqr#1#2{{\vcenter{\vbox{\hrule height.#2pt
 \hbox{\vrule width.#2pt height#1pt \kern#1pt
 \vrule width.#2pt}\hrule height.#2pt}}}}

\def\r{\rho}
\def\dd{\delta}

\def\hh{\hat{h}}
\def\hf{\hat{f}}
\def\hg{\hat{g}}
\def\hK{\hat{K}}
\def\aa1{\phi}
\def\cc1{\psi}
\def\hh{\hat{h}}

\def\D{\Delta}
\def\k{\kappa}
\def\dilog{\rm dilog}
\def\ta{\tilde{a}}
\def\l{\lambda}
\def\z{\zeta}

\def\ha{\hat{a}}

\catcode`\@=12

\begin{document}


\title{The black hole in the throat -- thermodynamics\\\medskip
of strongly coupled cascading gauge theories}

\pubnum{%
UWO-TH-07/10\\ WIS/07/07-JUNE-DPP\\ SLAC-PUB-12552}
\date{June 2007}

\author{
Ofer Aharony$ ^{1,2}$, Alex Buchel$ ^{3,4}$ and  Patrick Kerner$^{3}$\\[0.4cm]
\it $^1$Department of Particle Physics, Weizmann Institute of Science,\\
\it Rehovot 76100, Israel\\[0.2cm]
\it $^2$SITP, Department of Physics and SLAC, Stanford University,\\
\it Stanford, California 94305, USA\\[0.2cm]
\it $ ^3$Department of Applied Mathematics\\
\it University of Western Ontario\\
\it London, Ontario N6A 5B7, Canada\\[0.2cm]
\it $ ^4$Perimeter Institute for Theoretical Physics\\
\it Waterloo, Ontario N2J 2W9, Canada\\[0.2cm]
}

\Abstract{We numerically construct black hole solutions
corresponding to the deconfined, chirally symmetric phase of
strongly coupled cascading gauge theories at various temperatures.
We compute the free energy as a function of the temperature, and we
show that it becomes positive below some critical temperature,
indicating the possibility of a first order phase transition at
which the theory deconfines and restores the chiral symmetry.}

\makepapertitle

\body

\version\versionno

\section{Introduction and Summary}

One of the most interesting outcomes of the AdS/CFT correspondence
\cite{ma1,gkp,Witten:1998qj,ma2} is the ability to study
quantitatively the deconfined phase of $3+1$ dimensional gauge
theories, something which cannot be done analytically for QCD
(except at temperatures much higher than the QCD scale). For
strongly coupled large $N$ gauge theories which have (at zero
temperature) a dual description given by a weakly curved string
background, the deconfined phase has a dual description in terms of
a black hole (black brane) background which can be reliably studied
in the supergravity approximation. Of course, theories with a weakly
curved dual are rather different from QCD in various ways, but one
can still hope that their deconfined phase will not behave that
differently from that of QCD, and in some cases these theories are
even continuously connected to (large $N$) QCD by varying a
dimensionless parameter, and one could hope that the dependence on
this parameter is small (at least at temperatures of order the
deconfinement temperature).

The simplest theory to study in this way, on which most of the
research thus far has focused, is the strong coupling limit of
${\cal N}=4$ supersymmetric Yang-Mills theory, whose deconfined
phase has a very simple description as a black hole in anti-de
Sitter space \cite{ma3}. This theory does not confine at low
temperatures, but its deconfined phase still seems to exhibit many
similarities to that of QCD. Obviously, it would be nice to have
additional examples of $3+1$ dimensional deconfined theories which
can be studied quantitatively by using their gravity dual, and, in
particular, examples of deconfined phases of confining theories,
in which one could study the dependence on the temperature
compared to the deconfinement scale\footnote{Note that in any
large $N$ gauge theory with a weakly curved string theory dual the
deconfinement transition is a first order phase transition. This
is similar to large $N$ QCD, but it is different from QCD itself,
so one should be careful when comparing the behavior of such
theories near the phase transition to that of QCD.}. So far there
is only one known example of such a deconfined background, which
is that corresponding to $4+1$ dimensional supersymmetric
Yang-Mills theory (with a specific UV completion) compactified on
a circle with anti-periodic boundary conditions for fermions
\cite{ma3,osy}. It would be nice to have additional examples,
especially since in the example above the physics at the
deconfinement scale is really five dimensional rather than four
dimensional.

In this paper we study the deconfined phase of the confining
``cascading gauge theories'' constructed by Klebanov and
collaborators \cite{kn,kt,ks}. The equations determining the
corresponding black hole solutions are quite complicated, and have
no known analytic solutions. At very high temperatures it is
possible to find analytic solutions in an expansion in inverse
powers of the logarithm of the temperature, and the leading order
solution in this expansion was found in \cite{bh2}. This solution
shows that at high temperatures the number of degrees of freedom in
the theory grows as the square of the logarithm of the temperature
\cite{bh0,run1,bh2}. In this paper we numerically solve the
equations for a wide range of values of the temperature, in the
supergravity approximation, and use the solutions to analyze the
thermodynamics of the deconfined phase\footnote{Similar numerical
solutions were studied in \cite{lpz}, but we do not understand the
parametrization used there to analyze the solutions. Presumably, our
solutions should be identical to (some of) the solutions of
\cite{lpz}, but our parametrization allows for a direct computation
of the thermodynamical properties of the solutions.}.

At low temperatures the ``cascading gauge theories'' spontaneously
break a discrete chiral symmetry (and also a continuous $U(1)_B$
symmetry) \cite{ks,Aharony:2000pp,Klebanov:2002gr,Gubser:2004qj}. At
high temperatures one expects these symmetries to be restored, but
apriori it is not obvious whether there is a single phase transition
from a low-temperature phase with confinement and chiral symmetry
breaking to a high-temperature phase with no confinement and chiral
symmetry restored, or whether additional phases also exist. In this
paper we only study deconfined phases in which the chiral symmetry
is restored; we plan to study the possibility of having other phases
in the future \cite{ab}. In the classical supergravity approximation
the free energy of the low temperature confined phase vanishes
(since it only arises at one-loop), and thus the phase transition to
a black hole background occurs at the lowest temperature for which
the free energy of a black hole background starts becoming negative.
Assuming that this transition goes directly to the chirally
symmetric black holes that we construct, we find that there is a
first order deconfinement transition at a temperature $T_{critical}
= 0.614111(3) \Lambda$, according to a specific definition of the
strong coupling scale $\Lambda$ that we describe in section
\ref{numerics}\footnote{It is easy to translate this definition to
other definitions of the strong coupling scale, such as the mass
gap. Note that, as in all theories with a gravity approximation, the
square root of the confining string tension is not a useful measure
of the strong coupling scale, since it must be much larger than all
other measures of this scale for gravity to be a good
approximation.}. The black hole backgrounds continue to exist also
at lower temperatures, but they have positive free energies so they
no longer dominate the thermodynamics. Presumably, as the
temperature is lowered further, the black hole backgrounds
eventually become singular; in this paper we only compute the
numerical solutions until a temperature slightly below the
deconfinement temperature, so we do not see this.

There are several interesting directions for further study. We are
currently working on checking whether the solutions we find are
stable to deformations which break the chiral symmetry, in order to
see if there are signs of a deconfined non-chirally-symmetric phase
appearing at intermediate temperatures \cite{ab}. The black hole
solutions that we find (numerically) can be used for a detailed
analysis of the properties of the deconfined phase (for instance its
hydrodynamical properties \cite{tranc} or jet quenching
\cite{jet1,jet2}); it would be interesting to understand the
similarities and differences between these properties and those of
deconfined conformal theories. It is possible to add flavors in
various ways to the ``cascading gauge theories''
\cite{fl1,fl3,fl6,fl5,fl2,fl4,flnew}, and it would be interesting to
study the flavor physics in the deconfined phase, and whether there
are any phase transitions associated with the flavor sector.

Our study is purely in the supergravity approximation; it would be
interesting to study the corrections to this approximation, in
particular those coming from string theory corrections to the
supergravity action. The cascading gauge theories have a continuous
dimensionless parameter such that in one extreme of this parameter
supergravity is a good approximation, while in the other extreme
they reduce to a standard ${\cal N}=1$ supersymmetric Yang-Mills
(SYM) theory. At zero temperature supersymmetry tells us that the
dependence on this parameter is smooth. However, it is not obvious
if the behavior of the deconfined phase is smooth as this parameter
in changed; in fact, it seems plausible \cite{ofer_talk} that as in
other similar cases \cite{osy} there would be a phase transition in
this phase, since the geometry in the supergravity regime does not
have any cycles shrinking at the horizon (except for the thermal
$S^1$), while in the SYM regime one expects the transition to be
independent of the KK modes so a two-cycle should still shrink (as
it does in the confined phase). Of course, even in the absence of
such a phase transition, the behavior of the theory could be
modified as the dimensionless parameter is changed, so our analysis
does not teach us directly about the phase structure of the large
$N$ ${\cal N}=1$ SYM theory.

This paper is organized as follows. In section 2 we describe our
ansatz for the black hole solutions and the equations of motion that
it leads to. In section 3 we describe the boundary conditions for
these equations. In section 4 we analyze the meaning of the
parameters appearing in these boundary conditions, and show how to
translate them into physical quantities such as the temperature and
free energy. In section 5 we describe our numerical procedure and
present the ``bare'' numerical results. In section 6 we translate
these results into physical properties, and present the results for
physical quantities such as the free energy and the expectation
values of various operators as a function of the temperature. An
appendix contains a perturbative analysis of the very high
temperature solutions; this is useful both in order to make sure
that our analysis is valid by verifying that it is consistent (at
least at very high temperatures) with the first law of
thermodynamics, and in order to test our numerical solutions at very
high temperatures by testing their agreement with the perturbative
expansion.

\section{The equations for the cascading black hole}

In this paper we compute the metrics corresponding to the finite
temperature behavior of the ``cascading gauge theory'' found in
\cite{kn,kt,ks}\footnote{See \cite{review} for a recent review of
this theory.}, which may be thought of as a specific $SU(K)\times
SU(K+M)$ ${\cal N}=1$ supersymmetric gauge theory, with a number of
colors $K$ which runs logarithmically with the energy scale
\cite{ks,bh0,run1,aby,aby1}. The ``cascading gauge theory'' has a
single dimensionless parameter (in addition to the integer $M$),
which in the gravitational description of this theory can be taken
to be $g_s M$ where $g_s$ is the string coupling (which is constant
in the zero temperature solution) and $M$ is the RR 3-form flux
(corresponding to the number of fractional branes). When this
parameter is large, the gravitational description of the background
is valid at all scales. On the other hand, when it is small, the
theory at low energies reduces exactly to the ${\cal N}=1$ SYM
theory, but the gravity dual is highly curved. We will only analyze
the theory in the regime of large $g_s M$, where the gravitational
approximation is good and all radii of curvature are large compared
to the string scale.

As in any other confining background, the low temperature behavior
of this theory is governed by a gas of hadrons; the gravity dual of
this description is simply given by a thermal identification ($t
\equiv t + 1/T$) of the zero temperature solution found in
\cite{ks}. As the temperature is increased one expects the theory to
deconfine; in the gravitational dual, deconfined phases are
described by black holes (whose horizon fills all of space, so they
are really black branes). Our goal in this paper will be to compute
the gravitational backgrounds corresponding to the deconfined phase
of the cascading gauge theory. Note that the low temperature phase
is stable all the way up to the Hagedorn temperature of the
confining theory (related to the confining string tension); when the
gravitational approximation is valid, this temperature is very large
compared to the characteristic mass scale of the gravitational
background (which determines the mass of the low-lying hadrons).
Thus, as in all other cases of confining backgrounds with gravity
duals, we expect the deconfinement transition to occur at a
temperature which is much smaller than the Hagedorn temperature,
which means that it should be a first order phase transition.

The cascading gauge theory has a $Z_{2M}$ chiral symmetry
\cite{ks,Klebanov:2002gr} which is spontaneously broken to $Z_2$ at
low temperatures (by gaugino condensation in the limit where the
theory is a pure SYM theory), and it has a $U(1)_B$ symmetry which
is also spontaneously broken \cite{Aharony:2000pp,Gubser:2004qj}. At
high enough temperatures we expect these symmetries to be restored
\cite{bh0,bh1}; this expectation is confirmed by the analysis of the
asymptotically high temperature black hole solutions in \cite{bh2}.
Apriori it is not obvious if the deconfinement transition happens
together with the global symmetry restoration transitions, or if the
transitions are separate. In this paper we will only look for
solutions which preserve the chiral symmetry and the $U(1)_B$
symmetry; the stability of these solutions with respect to
chiral-symmetry-breaking deformations will be analyzed in \cite{ab}.
We also assume that the solutions preserve the $SU(2)\times SU(2)$
global symmetry of the theory, which is preserved also at low
temperatures so it is reasonable to assume that it is preserved at
all temperatures.

The form of the gravitational background of the cascading gauge
theory at large radial variables (close to the boundary) was found
in \cite{kt} and is known as the Klebanov-Tseytlin solution; this
form preserves the full global symmetry. The solution at any
temperature is expected to asymptote to this background near the
boundary (the ``UV region''). In the zero temperature solution
\cite{ks} the $Z_{2M}$ and $U(1)_B$ symmetries are broken far from
the boundary (in the ``IR region''), but we will look for solutions
where they are preserved. Recall that the solution for $M=0$ (where
the theory does not cascade) is \cite{kw} $AdS_5\times T^{1,1}$, and
that the solution of \cite{kt} has a similar form but with the radii
of curvature (and the flux) varying logarithmically in the radial
coordinate. Since we are looking for solutions that preserve the
full global symmetry, we can perform a Kaluza-Klein reduction on the
$T^{1,1}$, and leave only the fields which are singlets of the
global symmetry group. In general there are 5 such
fields\footnote{We only consider the fields which are turned on in
the solutions that we are interested in; other fields, such as the
type IIB axion, are consistently set to zero.}; the five dimensional
graviton and 4 scalar fields. In the $M=0$ limit the scalar fields
have scaling dimensions $\Delta=4,4,6,8$. The scalar fields are
various linear combinations of the dilaton, the overall volume of
the $T^{1,1}$, the relative size of the circle in $T^{1,1}$
(thinking of $T^{1,1}$ as a circle fibration over two 2-spheres),
and one mode coming from the RR fields.

We are looking for black hole solutions that preserve spatial
rotational and translational invariance, as well as time translation
invariance, so we can always choose a form of the five dimensional
metric where only 3 components are not vanishing -- $G_{00}$,
$G_{ii}$ ($i=1,2,3$) and $G_{rr}$ (where $r$ is the radial
position). We can use the freedom of reparametrizing the radial
coordinate to eliminate one of these degrees of freedom -- we will
choose our radial coordinate $x$ to be defined by
\be \eqlabel{xfixing} \frac{G_{00}}{G_{ii}} = -(1-x)^2 \ee
(with no summation over $i$). This choice is convenient since at
the boundary we expect the metric to be Lorentz-invariant so $x\to
0$, while at the horizon $G_{00}$ vanishes so $x\to 1$; the range
of the radial coordinate in our parametrization is thus $x \in
(0,1)$. Note that in the conformal $M=0$ case there is a simple
form of the black hole solution in this parametrization, which is
given by
\begin{equation} \eqlabel{pzerometric}
ds_{10}^2=\xi^2\ (2x-x^2)^{-1/2}\left(-(1-x)^2
dt^2+dx_1^2+dx_2^2+dx_3^2\right)+\frac{dx^2}{4(2x-x^2)^2}+\left(dT^{1,1}\right)^2,
\end{equation}
where $\left(dT^{1,1}\right)^2$  is the metric on $T^{1,1}$, and the
constant $\xi$ is related to the temperature $T$ as follows:
\begin{equation}
\xi=\pi T. \eqlabel{xiT}
\end{equation}

Motivated by the form of \eqref{pzerometric}, we write down the most
general ansatz for a black hole metric preserving all the symmetries
as\footnote{The frames $\{e_{\theta_a},e_{\phi_a}\}$ are defined as
in \cite{aby}, such that the metric on a unit size $T^{1,1}$ is
given by $\left(e_\psi^2\right)+ \sum_{a=1}^2
\left(e_{\theta_a}^2+e_{\phi_a}^2\right)$.} :
\begin{equation}
\begin{split}
ds_{10}^2=&h^{-1/2}(2x-x^2)^{-1/2}\left(-(1-x)^2 dt^2+dx_1^2+dx_2^2+dx_3^2\right)+G_{xx}(dx)^2\\
&+h^{1/2} [f_2\ \left(e_\psi^2\right)+ f_3\ \sum_{a=1}^2
\left(e_{\theta_a}^2+e_{\phi_a}^2\right)],
\end{split}
\eqlabel{ktm}
\end{equation}
where $h$, $f_2$ and $f_3$ are some functions of the radial
coordinate $x$. There is also a dilaton $g(x)$, and form fields
given by
\begin{equation}
\begin{split}
&F_3=P\ e_\psi \wedge \left(e_{\theta_1}\wedge e_{\phi_1}-e_{\theta_2}\wedge e_{\phi_2}\right)\,,\qquad
B_2=\frac{K}{2 P}\ \left(e_{\theta_1}\wedge e_{\phi_1}-e_{\theta_2}\wedge e_{\phi_2}\right),\\
&F_5=\calf_5+\star \calf_5\,,\qquad \calf_5=-K\ e_\psi\wedge e_{\theta_1}\wedge e_{\phi_1}\wedge e_{\theta_2}\wedge e_{\phi_2}\,,
\end{split}
\eqlabel{forms}
\end{equation}
where $K$ is a function of the radial coordinate $x$. The constant
$P$ appearing in \eqref{forms} is a constant times the quantized
flux $M$; we will write down the precise constant in terms of the
five dimensional Newton's constant below. We will find it simpler
to work in terms of $P$ rather than $M$, and we will only go back
to using the integer flux $M$ in the final section. After we
gauge-fixed the radial coordinate by \eqref{xfixing}, we have a
constraint equation coming from the equation of motion of this
variable; we can use this equation to solve for $G_{xx}$, which is
given by
\begin{equation}
\begin{split}
G_{xx}=&\frac{\sqrt{h} f_3^2}{ 2 (x-1) P^2 g^2 (2-x)^2 x^2 \D }
\biggl(12 P^2 f_3^2 g^2 f_2 h^2 (1-x)\\&+f_2 x^2 (2 P^2 f_3^2 g^2
h'^2 -12 P^2 g^2 h^2 f_3'^2+K'^2 h g
+2 P^2 h^2 f_3^2 g'^2) (x-1) (2-x)^2\\&-4 x P^2 f_3 g^2 h f_2 (2-x) (x^2-2 x+2) (h' f_3+4 f_3' h)\\
&+4 x P^2 f_3 (2-x) g^2 h^2 (2 x f_3' (1-x) (2-x)-(x^2-2 x+2) f_3)
f_2'\biggr),
\end{split}
\eqlabel{forgxx}
\end{equation}
with
\begin{equation}
\D\equiv K^2+8 h^2 f_3^2 f_2 (f_2-6 f_3)+2 h f_3^2 P^2 g.
\eqlabel{Deltadef}
\end{equation}

All in all, we have 5 scalar functions of $x$ that we need to solve
for : $h$, $f_2$, $f_3$, $g$ and $K$. We can derive the equations of
motion for these fields, in the supergravity approximation, either
directly from the ten dimensional type IIB supergravity action, or
by first reducing this action to five dimensions and then deriving
the equations of motion. The equations that we find take the
following rather complicated form :

\begin{equation}
\begin{split}
0=&h''-[8 h f_2 (f_2-6 f_3)+g P^2]\frac{ f_3^2 h'^2}{\D} +[8 x h^2 f_3^2 f_2 (f_2-6 f_3) (x-2)\\
&+K^2 (3 x^2-6 x+4)+4 h f_3^2 P^2 g (1-x)^2]\frac{h'}{x (1-x) (2-x) \D} \\
&-6  (K^2+h f_3^2 P^2 g)\frac{h f_3'^2}{f_3^2 \D }
+[8 h^2 f_3^2 f_2 (f_2-6 f_3)+3 K^2+4 h f_3^2 P^2 g]\frac{K'^2}{4g f_3^2 P^2 \D }\\
&+(K^2+h f_3^2 P^2 g) \frac{h g'^2}{g^2 \D }
-2 h (K^2+h f_3^2 P^2 g) [2 x f_3' (1-x) (2-x)-(x^2-2 x+2) f_3]\\
&\times\frac{f_2'}{f_3 f_2 x (1-x) (2-x) \D }
+ 8 (x^2-2 x+2) (K^2+h f_3^2 P^2 g) \frac{h f_3'}{x (1-x) (2-x) f_3 \D }\\
&-2  [7 h f_3^2 P^2 g+16 h^2 f_3^2 f_2 (f_2-6 f_3)+5
K^2]\frac{h}{(2-x)^2 x^2 \D },
\end{split}
\eqlabel{eom1}
\end{equation}
\begin{equation}
\begin{split}
0=&f_2''-\frac{f_2'^2}{f_2}
-[4 x f_3' f_3 h (1-x) (2-x) (g P^2-8 f_2^2 h)+8 f_2 h^2 f_3^2 \{6 x f_3 (x-2)\\
&+f_2 (4-2 x+x^2)\} +x K^2 (2-x)-4 h f_3^2 P^2 g
(1-x)^2]\frac{f_2'}{(1-x) (2-x) x \D }
\\
&+f_3^2 (g P^2-8 f_2^2 h) \frac{f_2h'^2}{h\D} -6 h (g P^2-8 f_2^2 h) \frac{f_2f_3'^2}{\D}
\\
&-(K^2+24 f_3^2 h^2 f_2 (f_2-2 f_3)) \frac{f_2K'^2}{4 h f_3^2 P^2 g \D }
+ h f_3^2 (g P^2-8 f_2^2 h) \frac{f_2g'^2}{g^2 \D }
\\
&+ 2 f_3^2 (x^2-2 x+2) (g P^2-8 f_2^2 h) \frac{f_2h'}{(2-x) (1-x) x \D }
\\
&+ 8 h f_3 (x^2-2 x+2) (g P^2-8 f_2^2 h) \frac{f_2f_3'}{(2-x) (1-x) x \D }
\\
&+2 [16 f_3^2 h^2 f_2 (2 f_2-3 f_3)+K^2-h f_3^2 P^2
g]\frac{f_2}{(x-2)^2 x^2 \D },
\end{split}
\eqlabel{eom2}
\end{equation}
\begin{equation}
\begin{split}
0=&f_3''-[2 P^2 g h f_3^2+8 f_3^2 f_2 h^2 (4 f_2-15
f_3)+K^2]\frac{f_3'^2}{f_3 \D } +4 (f_2-3 f_3) \frac{f_3^3
f_2h'^2}{\D}
\\
&+2 (f_2-3 f_3) \frac{f_3 h f_2K'^2}{P^2 g \D }
+ 4 (f_2-3 f_3) \frac{f_3^3 h^2 f_2g'^2}{g^2 \D}
\\
&+8 (x^2-2 x+2) (f_2-3 f_3) \frac{h f_3^3 f_2}{(2-x) (1-x) x \D } h'
\\
&-8 (f_2-3 f_3) [2 x f_3' (1-x) (2-x)-(x^2-2 x+2) f_3] \frac{h^2 f_3^2f_2'}{(2-x) (1-x) x \D }\\
&+\{8 f_2 f_3^2 [(5 x^2-10 x+8) f_2-6 f_3 (3 x^2-6 x+4)] h^2+2 x P^2 g f_3^2 (x-2) h\\
&+x K^2 (x-2)\}\frac{f_3'}{(1-x) x (2-x) \D}
+[4 P^2 g h f_3^2+2 K^2-8 f_3^2 f_2 (3 f_3+f_2) h^2]\\
&\times\frac{f_3}{x^2 (2-x)^2 \D},
\end{split}
\eqlabel{eom3}
\end{equation}
\begin{equation}
\begin{split}
0=&K''-\frac{KK'^2}{\D}+[h g f_3+2 h g f_3' (1-x)+h f_3 g' (1-x)+g
f_3 h' (1-x)]\frac{K'} {g h f_3 (x-1)}
\\
&+12 P^2 \frac{g K hf_3'^2}{\D}-2 \frac{g P^2 K f_3^2h'^2}{h \D }
 -2 P^2 \frac{K h f_3^2g'^2}{g \D }
+4 [2 x f_3' (1-x) (2-x)\\
&-(x^2-2 x+2) f_3] \frac{K g f_3 P^2 hf_2'}{f_2 (1-x) (2-x) x \D }
-4 P^2 g K (x^2-2 x+2) \frac{f_3^2h'}{(2-x) (1-x) x \D }\\
&-16 (x^2-2 x+2) \frac{K f_3 P^2 g hf_3'}{(2-x) (1-x) x \D }+12
\frac{f_3^2 h K P^2 g}{x^2 (2-x)^2 \D},
\end{split}
\eqlabel{eom4}
\end{equation}
\begin{equation}
\begin{split}
0=&g''-\frac{g'}{1-x}-[8 h^2 f_3^2 f_2 (f_2-6 f_3)+3 P^2 g h
f_3^2+K^2]\frac{g'^2}{g \D } -P^2 \frac{f_3^2 g^2h'^2}{h \D } +6 P^2
\frac{h g^2f_3'^2}{\D}
\\
&+(8 h^2 f_3^2 f_2 (f_2-6 f_3)+K^2)\frac{K'^2}{4P^2 f_3^2 h \D} -2
f_3^2 (x^2-2 x+2) P^2 \frac{g^2h'}{(2-x) (1-x) x \D} \\
&+2 f_3 g^2 P^2 [2 x f_3' (1-x) (2-x) -(x^2-2 x+2) f_3]
\frac{hf_2'}{f_2 (2-x) (1-x) x \D }
\\
&-8 (x^2-2 x+2) P^2 \frac{g^2 h f_3f_3'}{(2-x) (1-x) x \D} +6 P^2
\frac{g^2 f_3^2 h}{x^2 (2-x)^2 \D}.
\end{split}
\eqlabel{eom5}
\end{equation}

\section{Boundary conditions}
\label{bc}

In order to solve the equations of motion \eqref{eom1}-\eqref{eom5}
we need to specify boundary conditions, both at the asymptotic
boundary and at the horizon. We will require that asymptotically the
solution should match onto the Klebanov-Tseytlin (KT) solution, and
that it should be regular near the horizon.

\subsection{The UV boundary conditions}

Near the boundary $x\to 0$ it is possible to solve the equations by
a power series in $x$ and $\ln(x)$, whose leading term gives the KT
solution. This expansion takes the general form :
\begin{equation}
\begin{split}
h=&h_{0,0}- \frac{P^2 g_0}{8a_0^2}\ \ln (x)+\sum_{n=1}^{\infty}\sum_{k=1}^{n} h_{n,k}\ x^{n/2}\ \ln^k (x), \\
f_2=&a_0+\sum_{n=1}^{\infty}\sum_{k=1}^{n} a_{n,k}\ x^{n/2}\ \ln^k (x), \\
f_3=&a_0+\sum_{n=1}^{\infty}\sum_{k=1}^{n} b_{n,k}\ x^{n/2}\ \ln^k (x), \\
K=&4 h_{0,0} a_0^2-\frac 12 P^2 g_0-\frac 12 P^2 g_0 \ln (x)+\sum_{n=1}^{\infty}\sum_{k=1}^{n} K_{n,k}\ x^{n/2}\ \ln^k (x),\\
g=&g_0+\sum_{n=1}^{\infty}\sum_{k=1}^{n} g_{n,k}\ x^{n/2}\ \ln^k
(x).
\end{split}
\eqlabel{genexp}
\end{equation}
Most of the coefficients appearing in this expansion are not
independent; the independent coefficients correspond either to
parameters of the cascading gauge theory or to vacuum expectation
values (VEVs) of the operators dual to the fields we are solving
for. In the KT case there are 3 asymptotic parameters, which we
choose to be $g_0$, $h_{0,0}$ and $a_0$. $g_0$ is related to the
dimensionless parameter of the cascading gauge theory, and we will
see that one combination of the other parameters is related to the
temperature and the other is related to the dynamical scale of the
cascading theory. Naively we would expect to have 5 parameters
related to VEVs, but in fact there is one relation between the VEVs
which is given by the conformal anomaly equation (see \cite{aby}),
so we are left with four parameters corresponding to VEVs, which we
choose to be $\{a_{2,0},g_{2,0},a_{3,0},a_{4,0}\}$. Note that a VEV
appearing at order $x^{n/2}$ corresponds to an operator which has
dimension $2n$ in the conformal limit of the theory.

Using these 7 parameters we can solve for the coefficients in
\eqref{genexp} to any order we wish. It turns out that there are no
non-zero coefficients at order $\calo(x^{1/2})$. The non-zero
coefficients at the following orders are : \nxt order $\calo(x)$:
\begin{equation}
h_{2,1}=-\frac{3g_0 a_{2,0}}{28a_0^3}\ P^2, \qquad
h_{2,0}=\left(\frac{5g_0
a_{2,0}}{28a_0^3}-\frac{g_0}{16a_0^2}-\frac{g_{2,0}}{16a_0^2}\right)\
P^2+\frac{3a_{2,0} h_{0,0}}{7a_0}, \eqlabel{h2}
\end{equation}
\begin{equation}
b_{2,0}=-\frac 17 a_{2,0}, \qquad g_{2,1}=\frac{6g_0
a_{2,0}}{7a_0},
\eqlabel{f32g2}
\end{equation}
\begin{equation}
K_{2,1}=-\frac{6g_0 a_{2,0}}{7a_0}\  P^2,\qquad
K_{2,0}=\left(\frac{g_0 a_{2,0}}{a_0}-\frac 14 g_0-\frac 12
g_{2,0}\right)\ P^2+\frac{24}{7} a_0 a_{2,0} h_{0,0}. \eqlabel{k2}
\end{equation}
\nxt order $\calo(x^{3/2})$:
\begin{equation}
h_{3,0}=\frac{g_0 a_{3,0}}{60 a_0^3}\ P^2,\qquad
\eqlabel{h30b30K30}
b_{3,0}=-\frac 14 a_{3,0},\qquad
K_{3,0}=\frac{g_0 a_{3,0}}{6 a_0}\ P^2.
\end{equation}
\nxt Order $\calo(x^2)$: using the notation $\delta\equiv 139 P^2
g_0-120 h_{0,0} a_0^2$, we have
\begin{equation}
\begin{split}
h_{4,3}=-\frac{3g_0 a_{2,0}^2}{196a_0^4}\ P^2\,,\qquad
a_{4,2}=-\frac{12a_{2,0}^2}{245a_0}\,,\qquad
b_{4,2}=&-\frac{12a_{2,0}^2}{245a_0}\,,
\end{split}
\eqlabel{h43}
\end{equation}
\begin{equation}
\begin{split}
h_{4,2}=&\frac{1}{\delta} \biggl\{ \left(-\frac{142637g_0^2 a_{2,0}^2}{3920a_0^4}-\frac{75g_0^2 a_{2,0}}{16a_0^3}
-\frac{5g_0^2}{16a_0^2}+\frac{75g_0^2 a_{4,0}}{8a_0^3}+\frac{5g_{2,0}^2}{16a_0^2}
+\frac{139g_0 a_{2,0} g_{2,0}}{28a_0^3}\right)\ P^4\\
&+\frac{1002g_0 h_{0,0} a_{2,0}^2}{49a_0^2}\ P^2 -\frac{1440h_{0,0}^2 a_{2,0}^2}{49} \biggr\}\,,
\end{split}
\eqlabel{h42}
\end{equation}
\begin{equation}
\begin{split}
h_{4,1}=&\frac{1}{\delta} \biggl\{
\left(\frac{95g_0^2}{32a_0^2}+\frac{230383g_0^2 a_{2,0}^2}{23520a_0^4}-\frac{139g_0 a_{2,0} g_{2,0}}{24a_0^3}
-\frac{35g_0^2 a_{4,0}}{16a_0^3}-\frac{7g_{2,0}^2}{96a_0^2}-\frac{1423g_0^2 a_{2,0}}{224a_0^3}\right)\ P^4\\
&+\biggl(\frac{136327g_0 h_{0,0} a_{2,0}^2}{245a_0^2}+\frac 52 g_0 h_{0,0}- \frac{5h_{0,0} g_{2,0}^2}{g_0}
-\frac{528h_{0,0} a_{2,0} g_{2,0}}{7a_0}
-\frac{150 g_0 a_{4,0} h_{0,0}}{a_0}\\
&+\frac{570g_0 h_{0,0} a_{2,0}}{7a_0}\biggr)\ P^2
+\frac{5160h_{0,0}^2 a_{2,0}^2}{49} +\frac{5760a_0^2 h_{0,0}^3 a_{2,0}^2}{49g_0 P^2}\biggr\}\,,
\end{split}
\eqlabel{h41}
\end{equation}
\begin{equation}
\begin{split}
h_{4,0}=&\frac{1}{\delta} \biggl\{ \biggl(\frac{31973g_0^2 a_{2,0}^2}{17640a_0^4}+\frac{11g_0^2 a_{4,0}}{4a_0^3}
+\frac{201g_0^2 a_{2,0}}{56a_0^3}-\frac{219g_0^2}{64a_0^2} -\frac{139g_0 g_{2,0}}{32a_0^2}
+\frac{695g_0 a_{2,0} g_{2,0}}{252a_0^3}
\\
&-\frac{335g_{2,0}^2}{288a_0^2}\biggr)\ P^4
+\biggl(
\frac{35g_0 a_{4,0} h_{0,0}}{2a_0}+\frac{67g_0 h_{0,0} a_{2,0}}{4a_0}-\frac{167g_0 h_{0,0}}{8}
-\frac{262231g_0 h_{0,0} a_{2,0}^2}{2940a_0^2}\\
&+\frac{28 h_{0,0} a_{2,0} g_{2,0}}{a_0}
+\frac{15}{4} g_{2,0} h_{0,0}+\frac{5h_{0,0} g_{2,0}^2}{3g_0} \biggr)\ P^2
+600 a_0 a_{4,0} h_{0,0}^2\\
&+\frac{2104a_0 h_{0,0}^2 a_{2,0} g_{2,0}}{7g_0}-\frac{535356h_{0,0}^2 a_{2,0}^2}{245}
-\frac{2280a_0 a_{2,0} h_{0,0}^2}{7} +\frac{20 a_0^2 g_{2,0}^2 h_{0,0}^2}{g_0^2}\\
&-\frac{49536a_0^2 h_{0,0}^3 a_{2,0}^2}{49g_0 P^2}\biggr\}\,,
\end{split}
\eqlabel{h40}
\end{equation}
\begin{equation}
\begin{split}
a_{4,1}=&\frac{1}{\delta} \biggl\{ \left(
-\frac{12463g_0 a_{2,0}^2}{70a_0}-\frac{15}{2} g_0 a_{2,0}-\frac 12 g_0 a_0+15 g_0 a_{4,0}+\frac {a_0 g_{2,0}^2}{2g_0}\right)\ P^2
\\
&+\frac{26688}{245} a_0 h_{0,0} a_{2,0}^2
+\frac{48a_0^2 h_{0,0} a_{2,0} g_{2,0}}{7g_0}-\frac{1152a_0^3 h_{0,0}^2 a_{2,0}^2}{49P^2 g_0}\biggr\}\,,
\end{split}
\eqlabel{a41}
\end{equation}
\begin{equation}
\begin{split}
b_{4,1}=&\frac{1}{\delta} \biggl\{
\left(-\frac{7177g_0 a_{2,0}^2}{490a_0}-\frac{15}{2} g_0 a_{2,0}-\frac 12 g_0 a_0+15 g_0 a_{4,0}+\frac{a_0 g_{2,0}^2}{2g_0}\right)\ P^2
-\frac{7872a_0 h_{0,0} a_{2,0}^2}{245} \\
&+\frac{48a_0^2 h_{0,0} a_{2,0} g_{2,0}}{7g_0}-\frac{1152a_0^3 h_{0,0}^2 a_{2,0}^2}{49P^2 g_0}\biggr\}\,,
\end{split}
\eqlabel{b41}
\end{equation}
\begin{equation}
\begin{split}
b_{4,0}=&\frac{1}{\delta} \biggl\{
\left(\frac{6366g_0 a_{2,0}^2}{245a_0}+6 g_0 a_0+\frac{74}{7} g_0 a_{2,0}-41 g_0 a_{4,0}-\frac{6 a_0 g_{2,0}^2}{g_0}\right)\ P^2
+\frac{126144a_0 h_{0,0} a_{2,0}^2}{245} \\
&-\frac{576a_0^2 h_{0,0} a_{2,0} g_{2,0}}{7g_0}
-120 a_0^2 a_{4,0} h_{0,0}+\frac{480 a_{2,0} h_{0,0} a_0^2}{7}+\frac{13824a_0^3 h_{0,0}^2 a_{2,0}^2}{49P^2 g_0}\biggr\}\,,
\end{split}
\eqlabel{b40}
\end{equation}
\begin{equation}
g_{4,2}=\frac{18g_0 a_{2,0}^2}{49a_0^2}\,,\qquad
g_{4,1}=\frac{36g_0 a_{2,0}^2}{49a_0^2}+\frac{3g_0
a_{2,0}}{7a_0}+\frac{6a_{2,0} g_{2,0}}{7a_0}\,,
\eqlabel{g41}
\end{equation}
\begin{equation}
\begin{split}
g_{4,0}=&\frac{1}{\delta} \biggl\{
\left(
\frac{214407g_0^2 a_{2,0}^2}{980a_0^2}+\frac{15g_0^2}{4} +\frac{3243g_0^2 a_{2,0}}{28a_0}-
\frac{225g_0^2 a_{4,0}}{2a_0}
+\frac{263g_{2,0}^2}{4} +\frac{139g_{2,0} g_0}{2} \right)\
 P^2\\
&+\frac{7200g_0 h_{0,0} a_{2,0}^2}{49} -\frac{360a_0 h_{0,0} a_{2,0} g_{2,0}}{7}
-\frac{360a_0 g_0 h_{0,0} a_{2,0}}{7} -60 a_0^2 g_{2,0} h_{0,0}\\
&-\frac{60 a_0^2 g_{2,0}^2 h_{0,0}}{g_0}
+\frac{8640a_0^2 h_{0,0}^2 a_{2,0}^2}{49P^2}\biggr\}\,,
\end{split}
\eqlabel{g40}
\end{equation}
\begin{equation}
K_{4,2}=-\frac{12g_0a_{2,0}}{35a_0^2}P^2\eqlabel{K42}\,,
\end{equation}
\begin{equation}
\begin{split}
K_{4,1}=&\frac{1}{\delta}\biggl\{\biggr(-\frac{4701g_0^2a_{2,0}}{56a_0}
-\frac{9719g_0^2a_{2,0}^2}{56a_0^2}+\frac{195a_0^2a_{4,0}}{4a_0}+\frac{13g_{2,0}^2}{8}
-\frac{13g_0^2}{8}\\
&-\frac{417g_0a_{2,0}g_{2,0}}{14a_0}\biggr)P^4
+\biggl(\frac{360g_0a_0a_{2,0}h_{0,0}}{7}+\frac{51096g_0h_{0,0}a_{2,0}^2}{245}\\
&+48a_0h_{0,0}a_{2,0}g_{2,0}\biggr)P^2-\frac{12384a_{2,0}^2a_0^2h_{0,0}^2}{49}\biggr\}\,,
\end{split}
\eqlabel{K41}
\end{equation}
\begin{equation}
\begin{split}
K_{4,0}=&\frac{1}{\delta}\biggl\{\biggl(-\frac{189g_0^2}{16}-\frac{123g_0^2a_{2,0}}{56a_0}
+\frac{97g_0^2a_{4,0}}{4a_0}+\frac{23885g_0^2a_{2,0}^2}{392a_0^2}+\frac{139g_0a_{2,0}g_{2,0}}{14a_0}
-\frac{139g_0g_{2,0}}{4}\\
&-\frac{57g_{2,0}^2}{4}\biggr)P^4+\biggl(\frac{45g_0a_0^2h_{0,0}}{2}
+\frac{283077g_0h_{0,0}a_{2,0}^2}{245}+30a_0^2g_{2,0}h_{0,0}-390g_0a_0a_{4,0}h_{0,0}\\
&+\frac{2973g_0a_0a_{2,0}h_{0,0}}{7}+\frac{240a_0h_{0,0}a_{2,0}g_{2,0}}{7}\biggr)P^2
-\frac{1440a_0^3h_{0,0}^2a_{2,0}g_{2,0}}{7g_0}\\
&-\frac{1440a_0^3h_{0,0}^2a_{2,0}}{7}
+\frac{35712a_{2,0}^2a_0^2h_{0,0}^2}{49}
\biggr\}\,.
\end{split}
\eqlabel{K40}
\end{equation}

\subsection{The IR boundary conditions}

Next, we discuss the behavior of solutions to
\eqref{eom1}-\eqref{eom5} near the horizon, $x\to 1$. Introducing a near-horizon
coordinate
\begin{equation}
y\equiv 1-x
\eqlabel{horcoor}
\end{equation}
we find that in order for the solutions \eqref{ktm} to have a
non-singular Schwarzschild horizon, the functions
$\{h,f_2,f_3,g,K\}$ must all be even functions of $y$ with a good
Taylor series expansion around $y=0$ :
\begin{equation}
\begin{split}
h=&\sum_{n=0}^\infty h_{n}^h\ y^{2n},\qquad f_2=\sum_{n=0}^\infty
a_{n}^h\ y^{2n}\,,\qquad
f_3=\sum_{n=0}^\infty b_n^h\ y^{2n}\,,\\
K=&\sum_{n=0}^\infty k_n^h\ y^{2n},\qquad g=\sum_{n=0}^\infty g_n^h\
y^{2n}\,.
\end{split}
\eqlabel{genexph}
\end{equation}
When one solves the equations of motion perturbatively in $y$, one
finds that the solutions are labeled by six independent parameters,
which one can choose to be
$\{h_0^h,a_0^h,b_0^h,k_0^h,g_0^h,a_1^h\}$.
Naively one might think that all five equations of motion
\eqref{eom1}-\eqref{eom5} would have one normalizable mode and one
non-normalizable mode near the horizon, so that requiring a regular
solution will set the coefficients of the non-normalizable modes to
zero and leave us with 5 parameters. However, it turns out that for
one combination of the equations both modes are normalizable near
the horizon, leading to the extra parameter; this is related to a
scaling symmetry of the geometry \eqref{ktm} which we will discuss
in the next subsection, which implies that one combination of
parameters is not determined before choosing a scale \footnote{Such
a scaling symmetry in geometries with translationally invariant
horizons was also noticed in \cite{bln2}. }.

Using these 6 parameters we can solve for the coefficients in
\eqref{genexph} to any order we wish. Using the notation
$\dd^h\equiv 8 h_0^h (a_0^h)^2-P^2 g_0^h$, we have \nxt order
$\calo(y^2)$:
\begin{equation}
\begin{split}
&h_1^h=\frac {1}{\dd^h}\left\{8 (a_0^h)^2 (h_0^h)^2-\frac{(k_0^h)^2}{2(b_0^h)^2} -\frac{a_1^h (k_0^h)^2}{(b_0^h)^2 a_0^h}-\left(\frac 32 h_0^h g_0^h+\frac{h_0^h g_0^h a_1^h}{a_0^h}\right) P^2\right\}\,,\\
&b_1^h=\frac {1}{\dd^h}\left\{\frac 12 g_0^h b_0^h P^2-2 h_0^h b_0^h \left(3 (a_0^h)^2-3 a_0^h b_0^h+2 a_0^h a_1^h-6 a_1^h b_0^h\right)
\right\}\,,\\
&k_1^h=\frac {1}{\dd^h}\left\{\frac{P^2 g_0^h k_0^h (a_0^h+2 a_1^h)}{a_0^h}\right\}\,,\qquad g_1^h=\frac {1}{\dd^h}\left\{ \frac{(g_0^h)^2 P^2 (a_0^h+2 a_1^h)}{2 a_0^h}\right\}\,.\\
\end{split}
\eqlabel{horc1}
\end{equation}
\nxt order $\calo(y^4)$:
\begin{equation}
\begin{split}
&k_2^h=\frac {1}{(\dd^h)^2}\left\{- \frac{g_0^h P^2 k_0^h (a_0^h+2 a_1^h)^2}{2(a_0^h)^2} \biggl(4 h_0^h (a_0^h)^2-12 h_0^h a_0^h b_0^h-g_0^h P^2\biggr)\right\}\,,\\
\end{split}
\eqlabel{horc5}
\end{equation}
\begin{equation}
\begin{split}
&g_2^h=\frac {1}{(\dd^h)^2}\left\{-\frac{(g_0^h)^2 P^2(a_0^h+2 a_1^h)^2}{4(a_0^h)^2}
\biggl( 2 h_0^h (a_0^h)^2-6 h_0^h a_0^h b_0^h-g_0^h P^2\biggr)\right\}\,,\\
\end{split}
\eqlabel{horc6}
\end{equation}
\begin{equation}
\begin{split}
&h_2^h=\frac {1}{(\dd^h)^2}\biggl\{\frac{4 h_0^h a_0^h}{(b_0^h)^2}
\left(16 (b_0^h)^2 (a_0^h)^3 (h_0^h)^2- a_0^h (k_0^h)^2-2 a_1^h
(k_0^h)^2\right)+\frac{h_0^h (g_0^h)^2 P^4 (3 a_0^h+2 a_1^h)
}{2a_0^h}
\\&-\frac {g_0^hP^2}{4(a_0^h)^2 (b_0^h)^2} \biggl(
78 (h_0^h)^2 (b_0^h)^2 (a_0^h)^4+6 (h_0^h)^2 (a_0^h)^3 (b_0^h)^3-8 (h_0^h)^2 (a_0^h)^2 (a_1^h)^2 (b_0^h)^2
\\&+24 (h_0^h)^2 (a_1^h)^2 (b_0^h)^3 a_0^h
+24 (h_0^h)^2 (a_0^h)^3 a_1^h (b_0^h)^2+24 (h_0^h)^2 (a_0^h)^2 a_1^h (b_0^h)^3-(k_0^h)^2 (a_0^h)^2\\
&+4 (k_0^h)^2 (a_1^h)^2\biggr)\biggr\}\,,\\
\end{split}
\eqlabel{horc2}
\end{equation}
\begin{equation}
\begin{split}
&a_2^h=\frac {1}{(\dd^h)^2}\biggl\{-12 (h_0^h)^2 a_0^5+12 b_0^h (h_0^h)^2 (a_0^h)^4+16 (h_0^h)^2 (a_0^h)^4 a_1^h+48 (h_0^h)^2 (a_0^h)^3 a_1^h b_0^h\\
&+48 (h_0^h)^2 (a_0^h)^3 (a_1^h)^2+48 (h_0^h)^2 (a_0^h)^2 b_0^h (a_1^h)^2+\biggl(\frac 72 (a_0^h)^3 h_0^h g_0^h-\frac 32 b_0^h g_0^h h_0^h (a_0^h)^2\\
&-2 h_0^h (a_0^h)^2 a_1^h g_0^h-6 h_0^h a_0^h a_1^h b_0^h g_0^h-10 h_0^h a_0^h (a_1^h)^2 g_0^h-6 h_0^h (a_1^h)^2 g_0^h b_0^h\biggr) P^2
\\&+\biggl(-\frac 38 a_0^h (g_0^h)^2-\frac 12 (g_0^h)^2 a_1^h\biggr) P^4\biggr\}\,,\\
\end{split}
\eqlabel{horc3}
\end{equation}
\begin{equation}
\begin{split}
&b_2^h=\frac {1}{(\dd^h)^2}\biggl\{(h_0^h)^2 (-45 b_0^h (a_0^h)^3+16 (a_0^h)^3 a_1^h+45 (b_0^h)^2 (a_0^h)^2-132 (a_0^h)^2 a_1^h b_0^h\\
&+180 (b_0^h)^2 a_0^h a_1^h-84 a_0^h b_0^h (a_1^h)^2+180 (b_0^h)^2 (a_1^h)^2) b_0^h+\frac{h_0^h g_0^hP^2b_0^h}{4a_0^h} \biggl(6 (a_0^h)^3+9 b_0^h (a_0^h)^2\\
&+12 a_0^h a_1^h b_0^h+8 a_0^h (a_1^h)^2-12 (a_1^h)^2 b_0^h\biggr)
-\frac 18 (g_0^h)^2 b_0^h P^4\biggr\}\,.\\
\end{split}
\eqlabel{horc4}
\end{equation}

\section{Mapping of parameters to the field theory}

\subsection{Translation to the parametrization of \cite{aby}}
\label{translate}

In this section we wish to understand the physical meaning of the 3
parameters which we used in the previous section to parameterize our
theory (in the UV) -- $h_{0,0}$, $a_0$ and $g_0$. As we mentioned,
$g_0 P$ is the dimensionless parameter of the cascading theory
(which must be large for the gravity approximation to be valid),
while $h_{0,0}$ and $a_0$ are related to the scale of cascading
theory and to the temperature. Note that our ansatz \eqref{ktm} is
invariant under a scaling symmetry taking
\begin{equation} (t,\vec{x}) \to \lambda^{-2}(t,\vec{x})\,,\qquad
h\to \l^{-2}\ h\,,\qquad f_2\to \l f_2\,,\qquad  f_3\to \l f_3\,,
\eqlabel{scaling}
\end{equation}
and leaving all other functions in our solution (as well as the
coordinate $x$) invariant. In terms of our asymptotic parameters,
this scaling transformation leaves $h_{0,0} a_0^2$ invariant,
meaning that this combination is a function of the dimensionless
parameter of our theory, which is the ratio between the temperature
and some scale $\Lambda$ which characterizes the cascading theory.
We can choose this scale $\Lambda$ to be, say, the mass of the
lightest glueball, or the square root of the string tension. We will
find it more convenient to use a different definition of $\Lambda$
which will be described below.

The parametrization \eqref{ktm} we used above for the solution
breaks down at zero temperature, since it assumes the existence of a
horizon. In order to understand which combinations of our parameters
depend on the temperature and which do not it is convenient to
switch to a different parametrization of the geometry, which is
valid also at low temperatures : an example of such a
parametrization is given by \cite{aby}
\begin{equation}
\begin{split}
ds_{10}^2=&\hh^{-1/2} \r^{-2}\left(-\hf^2 dt^2+dx_1^2+dx_2^2+dx_3^2\right)+\hh^{1/2}\r^{-2}(d\r)^2\\
&+\hh^{1/2} \hf_2\ \left(e_\psi^2\right)+ \hh^{1/2} \hf_3\
\sum_{a=1}^2 \left(e_{\theta_a}^2+e_{\phi_a}^2\right),
\end{split}
\eqlabel{ktr}
\end{equation}
where $\{\hh,\hf_2,\hf_3,\hf\}$ are functions of $\r$. In this
parametrization the supersymmetric zero temperature solution is
characterized by two parameters: the value of the string coupling
$\hg_0$, and the coefficient of the warp factor $\hh_{0,0}$; in
terms of these parameters we can write the asymptotic solution for
small $\rho$ as
\begin{equation}
\begin{split}
&\hh=\hh_{0,0}-\frac 12 \hg_0 P^2\ln(\r)\,,\qquad \hK=4\hh_{0,0}-\frac 12 \hg_0 P^2 -2 \hg_0 P^2\ln (\r)\,,\\
&\hg=\hg_0,\qquad   \hf=\hf_2=\hf_3=1\,.
\end{split}
\eqlabel{susy}
\end{equation}
Note that the ansatz \eqref{ktr} is invariant under a joint
rescaling of the $x,t$ coordinates and the $\rho$ coordinate; such a
rescaling leads to a constant shift in $\hh_{0,0}$. Thus, we can
think of $\hh_{0,0}$ as determining the scale of the cascading
theory; note that this is independent of the temperature, since in
the parametrization \eqref{ktr} all IR effects (including the
effects of the temperature) are suppressed by powers of $\rho$.

We would like to match \eqref{susy} with the asymptotic solution
\eqref{genexp} used above. We require that as $\r\to 0$ (and
correspondingly $x\to 0$) all the corresponding warp factors in the
metric should agree to leading order, \ie,
\begin{equation}
\begin{split}
&\lim_{\{\r,x\}\to0}\frac{\r^{-2}\hf(\r)^2\hh(\r)^{-1/2}}{(1-x)^2(2x-x^2)^{-1/2}h(x)^{-1/2}}=1\,,\qquad
\lim_{\{\r,x\}\to0}\frac{\hh(\r)^{1/2}\hf_2(\r)}{h(x)^{1/2}f_2(x)}=1\,,\\
&\lim_{\{\r,x\}\to0}\frac{\hh(\r)^{1/2}\hf_3(\r)}{h(x)^{1/2}f_3(x)}=1\,,\qquad
\lim_{\{\r,x\}\to0} \frac{\hg(\r)}{g(x)}=1\,,\qquad
\lim_{\{\r,x\}\to0}\frac{\hK(\r)}{K(x)}=1\,.
\end{split}
\eqlabel{match}
\end{equation}
This matching uniquely identifies:
\begin{equation}
x = \frac 12 a_0^2\r^4 + {\rm higher\ orders}\,,\qquad
g_0=\hg_0\,,\qquad h_{0,0}\, a_0^2 = \hh_{0,0}+\frac 18 P^2 \hg_0\
\ln (\frac{a_0^2}{2})\,.
\eqlabel{match1}
\end{equation}

\subsection{Expectation values in the black hole background}

In order to proceed, we would like to compute the expectation values
of various operators in the cascading theories in terms of our
parameters; in particular we want to compute the expectation value
of the stress-energy tensor and of scalar operators which have
dimension four when $P\to 0$. The expectation values of these
operators were evaluated in \cite{aby} using the coefficients
appearing in the expansion in powers of $\rho$ of the functions
appearing in \eqref{ktr}, up to order $\rho^4$. In order to recycle
those results we need to translate our boundary expansion of the
previous section to the one of \cite{aby}, namely to write the
leading terms of the expansion of \cite{aby} in terms of our
parameters $\{h_{0,0},a_0,g_0,a_{2,0}\}$, as we did for the zeroth
order terms in \eqref{match1} \footnote{Note that the parameters
$\{a_{3,0}, a_{4,0}\}$ only show up at orders $\calo(\r^6)$ and
$\calo(\r^8)$, respectively, so they do not affect the expectation
values of these operators. }.

When we do the matching we have some freedom, since in the
parametrization \eqref{ktr} there is a freedom of performing
diffeomorphisms of $\rho$ depending on higher powers of $\rho$, that
only affect the higher order terms in the expansion. Of course this
freedom does not affect the eventual expectation values. We will fix
this freedom by making an explicit choice for $x$ as a function of
$\rho$ to order $\calo(\r^8)$, given by :
\begin{equation}
x = \frac 12\r^4 a_0^2-\r^8 a_0^3\left(\frac{5}{24} a_0+\frac{1}{14}
a_{2,0}\right)+\calo(\r^{10})\,. \eqlabel{frdef}
\end{equation}
We can now identify all the terms in the expansion of \cite{aby}
using our expansion of the previous section. Translating
\eqref{match1} to the notation of \cite{aby}, we find at order
$\r^{0}$
\begin{equation}
\begin{split}
&p_0=g_0\,,\qquad K_0=4 h_{0,0} a_0^2-\frac 12 P^2 g_0-\frac 12 P^2
g_0\ \ln(\frac{a_0^2}{2})\,,
\\
&G_{ij}^{(0)}=\eta_{ij}={\rm diag}\left(-1,1,1,1\right)\,.
\end{split}
\eqlabel{idenmatch0}
\end{equation}
All the coefficients in \cite{aby} at order $\r^{2}$ vanish, while
the independent parameters appearing at order $\r^4$ are given by
(again in the notation of \cite{aby})
\begin{equation}
\begin{split}
&a^{(4,1)}=0\,,\qquad a^{(4,2)}=0\,,\qquad a^{(4,3)}=0\,,\qquad
G_{tt}^{(4,0)}=a_0^2\,,\qquad G_{x_i x_i}^{(4,0)}=0\,,\\
&p^{(4,0)}=\frac{a_0^2g_{2,0}}{2g_0}+\frac 37 a_0 a_{2,0}\
\ln(\frac{a_0^2}{2})\,,\qquad a^{(4,0)}=\frac 47 a_0 a_{2,0}+\frac
13 a_0^2\,,\qquad b^{(4,0)}=\frac 13 a_0^2\,.
\end{split}
\eqlabel{idenmatch4}
\end{equation}

The one-point function of the stress energy tensor is given by \cite{aby}
\begin{equation}
8\pi G_5\langle T_{ij}\rangle=-\frac 12 G_{ij}^{(0)}\
G_a^{(4,0)a}+2G_{ij}^{(4,0)}+\frac 32 G_{ij}^{(0)} \
\left(b^{(4,0)}-a^{(4,0)}\right)\,, \eqlabel{stressvev}
\end{equation}
where $G_5$ is the five dimensional Newton's constant obtained
after we do the dimensional reduction on $T^{1,1}$ (see
\cite{aby}). In the normalizations that we are using, $G_5$ is
related to the ratio $P/M$ (using the careful analysis of
\cite{Herzog:2001xk}) by $G_5 = 8 \pi^3 P^4 / 81 M^4$. Using
\eqref{stressvev} we obtain that the energy density $\cale$ and
the pressure $\calp$ are given in terms of our parameters by
\begin{equation}
\begin{split}
&\cale\equiv \langle T_{tt}\rangle=\frac{1}{8\pi G_5}\ \left(\frac
32 a_0^2+\frac 67 a_0 a_{2,0}\right)\,, \\
&\calp\equiv \langle T_{x_i x_i}\rangle=\frac{1}{8\pi G_5}\
\left(\frac 12 a_0^2-\frac 67 a_0 a_{2,0}\right)\,.
\end{split}
\eqlabel{epressure}
\end{equation}
Since we do not have any chemical potentials, the free energy density $\calf$ is
\begin{equation}
\calf=-\calp=\frac{1}{8\pi G_5}\ \left(\frac 67 a_0 a_{2,0}-\frac
12 a_0^2\right)\,. \eqlabel{fenergy}
\end{equation}
The expectation values of the remaining two scalar operators which
have dimension 4 when $P=0$ are (using their normalization defined
in \cite{aby})
\begin{equation}
\begin{split}
&\langle \calo_{K_0}\rangle=\frac{24 a_0 a_{2,0}}{7 P^2 g_0}\,,\\
&\langle \calo_{p_0}\rangle=2 \frac{a_0^2g_{2,0}}{g_0^2}
+\frac{12a_0a_{2,0}}{7g_0}\ \left(1+\ln(\frac{a_0^2}{2})\right)\,.
\end{split}
\eqlabel{kpvev}
\end{equation}
Note that in general curved backgrounds there was an ambiguity in
some of the one-point correlation functions computed in \cite{aby},
but there is no such ambiguity when the asymptotic four dimensional
metric is flat (as in our case).

\subsection{The basic thermodynamic relation}

The equations above tell us, using the asymptotic values of the
fields, that
\begin{equation}
s T = \cale - \calf = \cale + \calp = \frac{a_0^2}{4 \pi G_5}
\eqlabel{bthermo}
\end{equation}
in the cascading background, where $s$ is the entropy density. On
the other hand, we can also compute the entropy density and
temperature directly at the horizon in terms of the horizon
parameters $\{h_0^h,a_0^h,b_0^h,k_0^h,g_0^h,a_1^h\}$ :
\begin{equation}
s=\frac{(a_0^h)^{1/2}(b_0^h)^2(h_0^h)^{1/2}}{4 G_5}\,,\qquad
T=\frac{1}{4\pi h_0^h b_0^h}\ \sqrt{\frac{2 (8 h_0^h
(a_0^h)^2-g_0^h P^2)}{a_0^h+2 a_1^h}}\,. \eqlabel{sT}
\end{equation}
At first sight the previous two equations seem to give a non-trivial
relation between some of our UV parameters and some of the IR
parameters related to the expansion near the horizon. However, it
turns out that $sT$ is a renormalization group flow invariant in
supergravity black brane geometries without a chemical potential
\cite{bl,bdkl}, so this relation is trivially satisfied in any
solution of our equations of motion.

To simplify notations we rewrite the metric \eqref{ktm} as
\begin{equation}
ds_{10}^2=-c_1^2\ dt^2+c_2^2\
\left(dx_1^2+dx_2^2+dx_3^3\right)+c_3^2\ \left(dx\right)^2+c_4^2\
\left( e_\psi^2\right)+c_5^2\
\sum_{a=1}^2\left(e_{\theta_a}^2+e_{\phi_a}^2\right)\,, \eqlabel{ktm2}
\end{equation}
where $c_i=c_i(x)$ can be identified by comparing \eqref{ktm} and
\eqref{ktm2}. Now, from the relation between components of the Ricci
tensor
\begin{equation}
R_{x_1}^{\ x_1}=R_t^{\ t}, \eqlabel{rel}
\end{equation}
we have a constraint\footnote{Equation \eqref{rel1} can also be
directly derived from \eqref{eom1}-\eqref{eom5}.}
\begin{equation}
\frac{c_2^4c_4c_5^4}{c_3}\ \left(\frac {c_1}{c_2}\right)'={\rm
constant}\,. \eqlabel{rel1}
\end{equation}
Evaluating the left-hand side of \eqref{rel1} near the horizon,
using the standard relations between the area of the horizon and the
entropy and between the surface gravity of the horizon and the
temperature, we have
\begin{equation}
\lim_{x\to 1_-}\ \frac{c_2^4c_4c_5^4}{c_3}\ \bigg|\left(\frac
{c_1}{c_2}\right)'\bigg|=8 \pi G_5\ sT\,. \eqlabel{rel2}
\end{equation}
On the other hand, evaluating the left-hand side of \eqref{rel1}
near the boundary and using the asymptotic solution \eqref{genexp}
we find
\begin{equation}
\lim_{x\to 0_+}\ \frac{c_2^4c_4c_5^4}{c_3}\ \bigg|\left(\frac
{c_1}{c_2}\right)'\bigg|=\lim_{x\to 0_+}\frac{h^{1/4}f_2^{1/2}f_3^2}
{(2x-x^2)G_{xx}^{1/2}}=2a_0^2\,. \eqlabel{rel3}
\end{equation}
Thus, our equation \eqref{bthermo} follows in a straightforward way
from the equations of motion.

\section{The numerical procedure}
\label{numerics}
\subsection{Reducing the number of parameters}

Before we begin the numerical solution of the equations, we can use
the symmetries of the problem to get rid of some of our parameters.
First, as mentioned above, the parameters $h_{0,0}$ and $a_0$ are
not scale invariant, but only the combination $h_{0,0} a_0^2$, which
is a function of the temperature divided by the dynamical scale. We
will choose as our parameter which is related to the dimensionless
temperature the combination $k_s$ defined by
\begin{equation}
P^2 g_0 k_s\equiv 4 h_{0,0} a_0^2-\frac 12 P^2 g_0\,. \eqlabel{num2}
\end{equation}
Equation \eqref{match1} now tells us that $[k_s - \ln(a_0^2/2) / 2]$
is independent of the temperature (it depends only on the dynamical
scale of the cascading theory). Thus, we can choose to define the
scale $\Lambda$ of this theory by a relation of the form
\begin{equation}
k_s \equiv \frac{1}{2} \ln\left(\frac{a_0^2}{\Lambda^4}\right) =
\frac{1}{2} \ln\left(\frac{4 \pi G_5 s T}{\Lambda^4}\right)\,.
\eqlabel{lambdadef}
\end{equation}
Using the expressions for the high temperature entropy density of
the theory computed in \cite{bh0,bh2}, we see that at high
temperatures $k_s\simeq (1/2) \ln(T^4/\Lambda^4)$, with corrections
scaling as $\ln(\ln(T/\Lambda))$. We will use $k_s$ instead of the
temperature as our basic dimensionless parameter, and use
\eqref{lambdadef} to translate between $k_s$ and $T/\Lambda$.

Having understood this relation, we can now use the scaling symmetry
\eqref{scaling} to set $a_0=1$, or, equivalently, use the fact that
the solution to the equations of motion depends on the 7 UV
parameters $\{g_0,h_{0,0},a_0,a_{2,0},g_{2,0},a_{3,0},a_{4,0}\}$
that we used in our expansion only through the six invariant
combinations
\begin{equation}
\biggl\{g_0\,, k_s\,, \ha_{2,0}\equiv \frac{a_{2,0}}{a_0}\,,
\ha_{3,0}\equiv \frac{a_{3,0}}{a_0}\,, \ha_{4,0}\equiv
\frac{a_{4,0}}{a_0} \,, g_{2,0}\biggr\}\,. \eqlabel{num5}
\end{equation}

Recall also that we are solving the theory in the supergravity
approximation, which includes only the leading order terms both in
the $g_s$ expansion and in the curvature ($\alpha'$) expansion. When
we neglect $g_s$ corrections, the action (and the equations of
motion we wrote) does not depend separately on $P^2$ and $g$ but
only on the combination $P^2 g$. We can thus set $g_0=1$, and recall
that whenever we have a factor of $P^2$ we really mean $P^2 g_0$.
Furthermore, when we neglect $\alpha'$ corrections, the action is
multiplied by a constant when we rescale the ten dimensional metric
by a constant factor (and rescale the $p$-forms accordingly), so
that the equations of motion are left invariant; this transformation
acts on our variables as
\begin{equation} \label{scaling2}
h\to \lambda^{-2} h\,,\qquad f_{2,3} \to \lambda^2 f_{2,3}\,,\qquad K\to
\lambda^2 K\,,\qquad g\to g\,,
\end{equation}
and it changes $P$ by $P\to \lambda P$. We can use this
transformation to relate the solutions for different values of $P$
(as long as we are in the supergravity approximation). Thus, we will
perform the numerical analysis for $P=1$, and we can use
\eqref{scaling2} to obtain the solutions for any other value of $P$.

As a test of our numerics, we can check if it reproduces the
solution at high temperatures which can be computed perturbatively
(as was done at leading order in \cite{bh2}, and at higher orders in the
appendix). This computation implies that the correct solution should
obey at large $k_s$ \eqref{a20a0}-\eqref{g20}:
\begin{equation}
\begin{split}
&\ha_{2,0}=\frac{7}{12}\ \frac{1}{k_s}-\frac{7}{24}\ \ln (2)\ \frac{1}{k_s^2}+\calo\left(k_s^{-3}\right),\\
&\ha_{3,0}=\frac{4}{5}\ \l_3^{[2]}\ \frac{1}{k_s}+\left(
\left(\frac{2}{15}-\frac {2}{5} \ln (2)\right) \l_3^{[2]}+\frac 45
\l_3^{[4]}\right)\ \frac{1}{k_s^2}+
\calo\left(k_s^{-3}\right)\,,\\
&\ha_{4,0}=\left(\frac{\ln (2)}{30}+\frac{1021}{1800}\right)\
\frac{1}{k_s}+\left( \frac{167809}{108000}-\frac{(\ln (2))^2}{360}
-\frac{781}{1200} \ln (2)+\eta_4^{[4]} \right)\frac{1}{k_s^2}
+\calo\left(k_s^{-3}\right)\,,\\
&g_{2,0}=\left(-\frac{1}{2}+\frac 12\ln (2)\right)
\frac{1}{k_s}+\left(\frac 14 \ln (2)-\frac 14 (\ln (2))^2
+\zeta_2^{[4]}\right)\ \frac{1}{k_s^2}+\calo\left(k_s^{-3}\right)\,,
\end{split}
\eqlabel{pert}
\end{equation}
where the values of the various constants appear in the appendix,
and we used the high-temperature relation between the value of $K$
at the horizon, which we denote by $K_\star$, and our dimensionless
parameter $k_s$:
\begin{equation}
K_\star=P^2\hg_0 \left(k_s+\frac 12 \ln
(2)+\calo\left(k_s^{-1}\right)\right)\,. \eqlabel{ksks}
\end{equation}

\subsection{Our numerical method}

As described above, for a given value of the temperature (or of
$k_s$) we have four parameters controlling the behavior of our
solutions near the UV. What we need to do is to find for which value
of these four parameters the solution is regular near the horizon,
and this will determine the correct vacuum expectation values for
this value of the temperature. The most naive way to proceed would
be to go over all possible values of these parameters, use these
values to determine the solution near the boundary, integrate the
equations of motion up to $x=1$, and see if the solution there is
regular or not. Unfortunately, we cannot integrate the equations
analytically but only numerically, and when we integrate the
equations near the horizon, numerical errors always generate modes
that blow up at the horizon, so we cannot really obtain solutions
that are regular at the horizon in this way.

One alternative might be to perform the integration in the opposite
direction -- start from a general solution near the horizon,
integrate the equations to the boundary, and see for which values of
our near-horizon parameters we find a regular solution at the
boundary (with the correct KT asymptotics). However, this suffers
from the same problem, that numerical errors generate modes that
grow near the boundary.

Thus, we are led to a procedure where we integrate the equations
both from the boundary and from the horizon towards the middle of
the interval $x=0.5$, and attempt to match a solution that we get by
integrating from the boundary with a solution that we get by
integrating from the horizon. After setting $g_0=1$, for a given
value of $k_s$, the UV behaviour \eqref{genexp} is determined by 4
parameters (related to operator VEVs) \eqref{num5}
\begin{equation}
\{ \ha_{2,0}\,,
\ha_{3,0}\,, \ha_{4,0}\,, g_{2,0}\}\,.
\eqlabel{uvpar}
\end{equation}
The IR behaviour \eqref{genexph} is determined by 6 horizon parameters
\begin{equation}
\{h_0^h,a_0^h,a_1^h,b_0^h,k_0^h,g_0^h\}\,.
\eqlabel{irrap}
\end{equation}
Matching a UV solution and an IR solution to
\eqref{eom1}-\eqref{eom5} at $x=0.5$ implies 10 constraints (5 for
matching the values of the functions, and 5 for matching their
derivative). Notice that we have precisely the same number of
constraints as necessary to uniquely determine all the UV and IR
parameters (\eqref{uvpar} and  \eqref{irrap}) for a given value of
$k_s$.

\begin{figure}[t]
\begin{center}
  \includegraphics[width=5in]{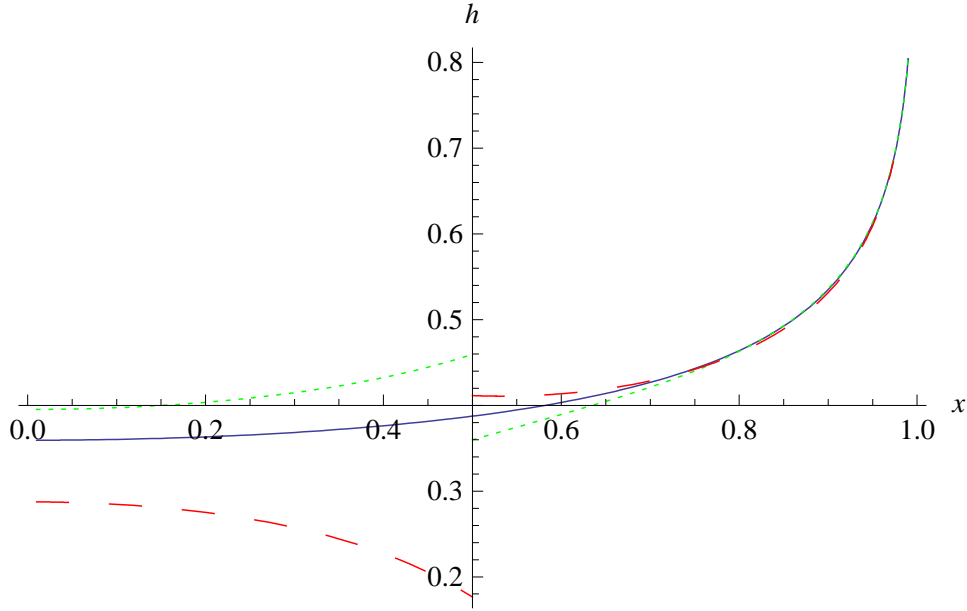}
\end{center}
  \caption{
Mismatch of $h_b(x)$ (for $x<0.5$) and $h_h(y\equiv 1-x)$ (for
$x>0.5$) for different values of the parameters, for $k_s=0.4$. The
solid (blue) curves correspond to ``correct'' values of the parameters
\eqref{uvpar} and \eqref{irrap}, with $||\vec{v}_{mismatch}||
\approx 9\times 10^{-6}$. The dotted (green) curves corresponding to all
values of parameters $10\%$ larger than the correct ones, produce
$||\vec{v}_{mismatch}||\approx 3\times 10^{-1}$. The dashed (red) curves
correspond to all values of parameters $20\%$ smaller than the
correct ones, giving $||\vec{v}_{mismatch}||\approx 8\times
10^{-1}$. } \label{fig1}
\end{figure}

Since both the boundary $x=0$ and the horizon $x=1$ are singular
points of the differential equations  \eqref{eom1}-\eqref{eom5},
we integrate the differential equations  \eqref{eom1}-\eqref{eom5}
from $x=0.01$ (for the boundary integration) and from $y=0.01$ (for
the horizon integration). In the former case the initial conditions
are specified by the asymptotic expansion \eqref{genexp} which we
developed to order $x^{9/2}$ (inclusive); in the case of the horizon
integration the initial conditions are specified by the asymptotic
expansion \eqref{genexph} to order $y^{10}$ (inclusive). The
coefficients of these asymptotic expansions generalize the results
presented in section \ref{bc}, and are available from the authors
upon request. The mismatch between the boundary and the horizon
integrations is encoded in the `mismatch vector'
$\vec{v}_{mismatch}$, defined by
\begin{equation}
\begin{split}
\vec{v}_{mismatch}=&\biggl(h_b-h_h,h_b'+h_h',f_{2,b}-f_{2,h},f_{2,b}'+f_{2,h}',f_{3,b}-f_{3,h},f_{3,b}'+f_{3,h}',\\
&K_b-K_h,K_b'+K_h',g_b-g_h,g_b'+g_h'\biggr)\bigg|_{x=y=0.5}\,,
\end{split}
\eqlabel{vmis}
\end{equation}
where the subscripts $ _h$ or $ _b$ correspond to functions
$\{h,f_2,f_3,K,g\}$ integrated from the horizon or boundary,
respectively, and the prime denotes derivatives with respect to $x$
or $y$. The UV parameters \eqref{uvpar} and the IR parameters
\eqref{irrap} are tuned to ensure that
\begin{equation}
||\vec{v}_{mismatch}|| < 10^{-5}\,. \eqlabel{norm}
\end{equation}
An illustration of the integration as a function of the UV and IR
parameters is presented in figure \ref{fig1}.

We performed the numerical integration using {\it Wolfram
Mathematica$ ^{\copyright}$6} with 40 digit precision, to ensure
sensitivity to the irrelevant operator parameters $\ha_{3,0}$ and
$\ha_{4,0}$.

\subsection{The numerical results}

We present the numerical results for the UV \eqref{uvpar} and the IR
\eqref{irrap} parameters as a function of $k_s$ in two
regimes\footnote{The IR parameters are presented only for small
values of $k_s$. Additional data are available from the authors upon
request.}: \nxt for large values of $k_s$, where we can check our
numerical results against the perturbative analytic predictions
\eqref{pert}; \nxt for an interval of small values of $k_s$  that
includes the first order transition point to a confined thermal
cascading background with broken chiral symmetry (as we will discuss
in the next section).

\subsubsection{Large values of $k_s$}

\begin{figure}[t]
 \hspace*{-20pt}
\psfrag{ks}{\raisebox{2ex}{\footnotesize\hspace{-0.8cm}$k_s$}}
\psfrag{a20}{\raisebox{0ex}{\footnotesize\hspace{0cm}$\ha_{2,0}$}}
\psfrag{a30}{\raisebox{0ex}{\footnotesize\hspace{0cm}$\ha_{3,0}$}}
  \includegraphics[width=3.0in]{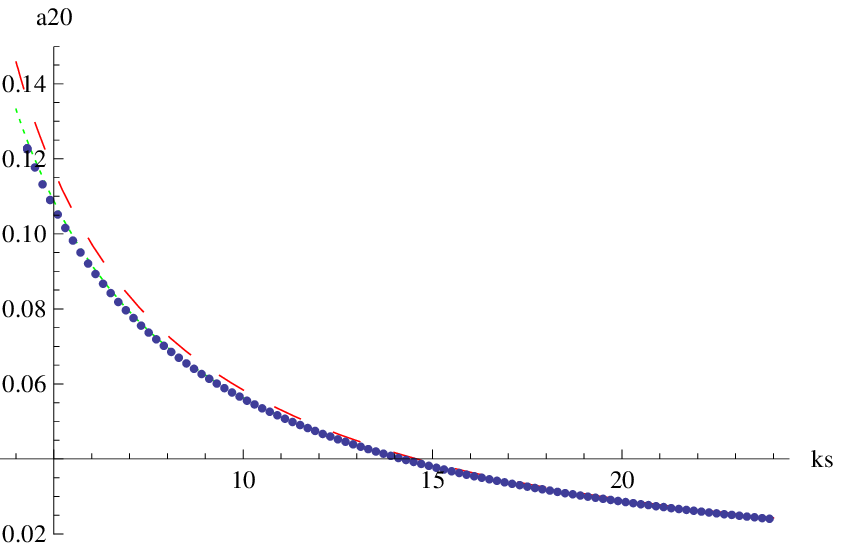}
  \includegraphics[width=3.0in]{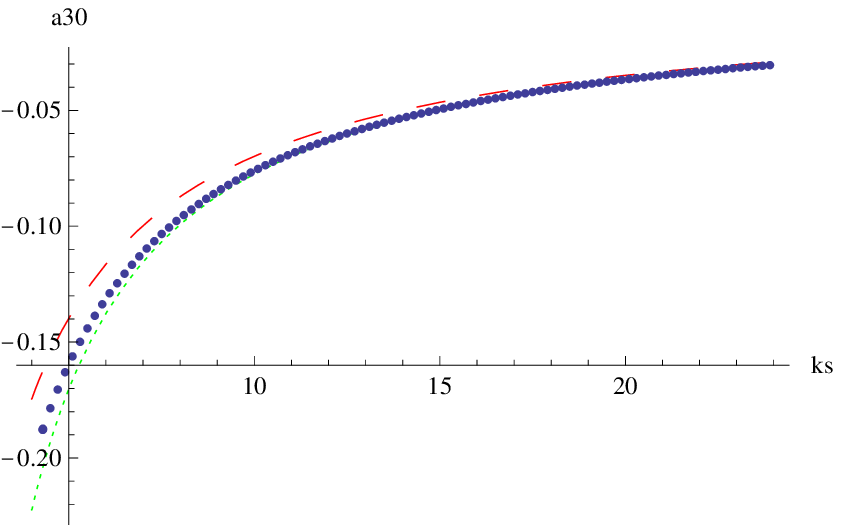}
  \caption{
Values of the UV parameters $\ha_{2,0}$ and $\ha_{3,0}$ as a
function of $k_s$ (blue points). The dashed/dotted (red/green)  curves
represent the perturbative $\calo(k_s^{-1})$/$\calo(k_s^{-2})$
asymptotics of the parameters, given by \eqref{pert}. } \label{fig2}
\end{figure}

\begin{figure}[t]
 \hspace*{-20pt}
\psfrag{ks}{\raisebox{2ex}{\footnotesize\hspace{-0.8cm}$k_s$}}
\psfrag{a40}{\raisebox{0ex}{\footnotesize\hspace{0cm}$\ha_{4,0}$}}
\psfrag{g20}{\raisebox{0ex}{\footnotesize\hspace{0cm}$g_{2,0}$}}
  \includegraphics[width=3.0in]{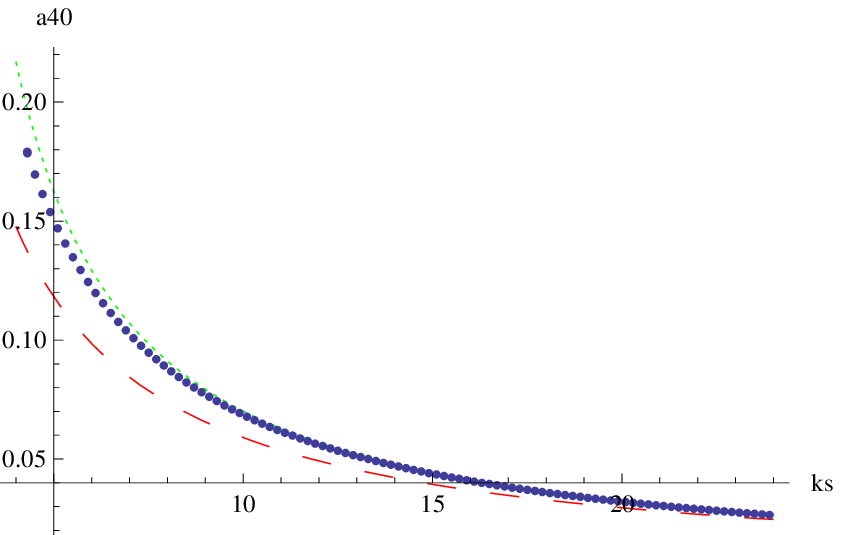}
  \includegraphics[width=3.0in]{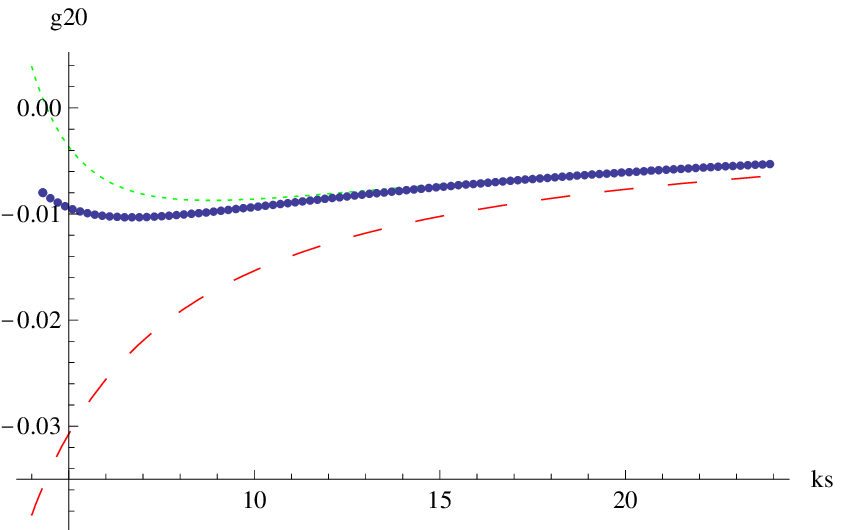}
  \caption{
Values of the UV parameters $\ha_{4,0}$ and $g_{2,0}$ as a function
of $k_s$ (blue points). The dashed/dotted (red/green) curves
represent the perturbative $\calo(k_s^{-1})$/$\calo(k_s^{-2})$
asymptotics of the parameters, given by \eqref{pert}. } \label{fig3}
\end{figure}

Figures \ref{fig2} and \ref{fig3} present the dependence of the UV
parameters $\{\ha_{2,0},\ha_{3,0},\ha_{4,0},g_{2,0}\}$ on $k_s\in
(4.29,24.0)$, with a step of $\Delta k_s=0.01$ (blue points). In
this regime the typical norm of the mismatch vector \eqref{vmis}
$||\vec{v}_{mismatch}||\sim 10^{-10}$ or less. The dashed and dotted
(red and green) curves represent the perturbative $\calo(k_s^{-1})$
and $\calo(k_s^{-2})$ asymptotics \eqref{pert}, respectively. Note
that the dotted (green) curves track our numerical data quite well
in this regime\footnote{In appendix A we confirm using the
perturbative high-temperature expansion that the cascading geometry
thermodynamics satisfies the first law of thermodynamics up to order
$\calo(k_s^{-3})$.}.

\subsubsection{Small values of $k_s$}

\begin{figure}[t]
 \hspace*{-20pt}
\psfrag{ks}{\raisebox{2ex}{\footnotesize\hspace{-0.8cm}$k_s$}}
\psfrag{a20}{\raisebox{0ex}{\footnotesize\hspace{0cm}$\ha_{2,0}$}}
\psfrag{a30}{\raisebox{0ex}{\footnotesize\hspace{0cm}$\ha_{3,0}$}}
\psfrag{kc}{\raisebox{2ex}{\footnotesize\hspace{0cm}$\left(k_{critical},\frac{7}{12}\right)$}}
  \includegraphics[width=3.0in]{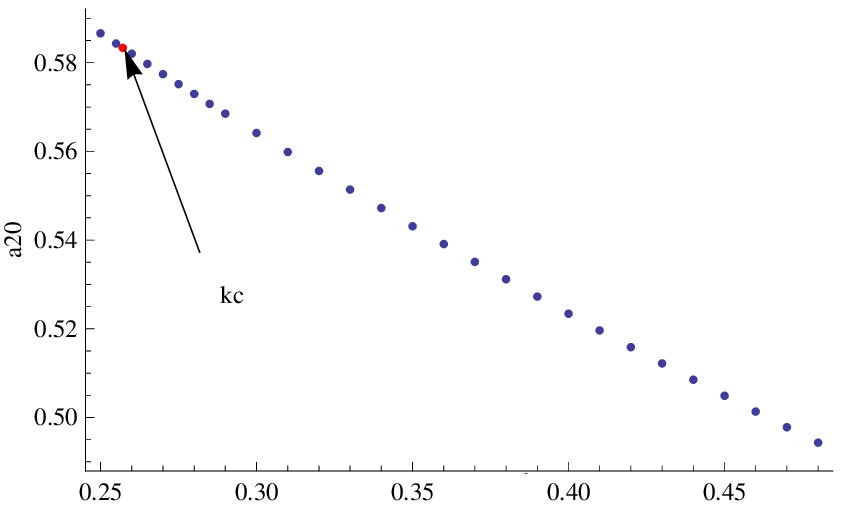}
  \includegraphics[width=3.0in]{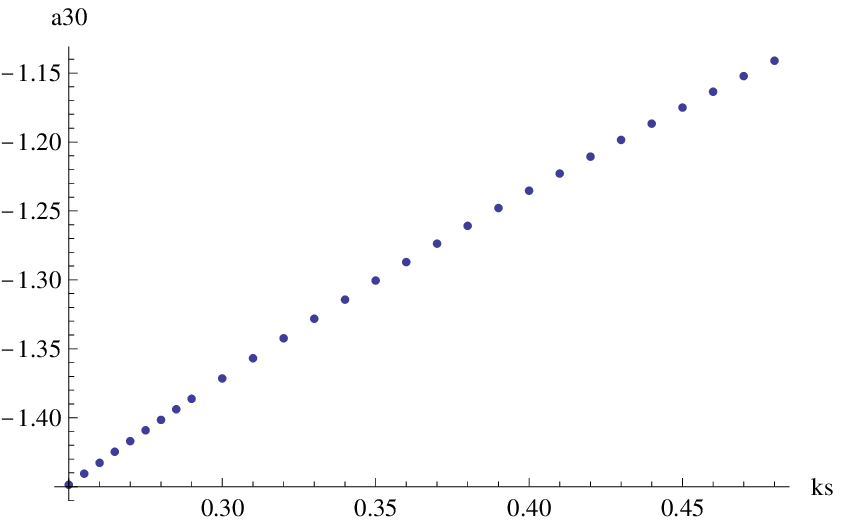}
  \caption{
The values of the UV parameters $\ha_{2,0}$ and $\ha_{3,0}$ as a
function of $k_s$.} \label{fig4}
\end{figure}

\begin{figure}[t]
 \hspace*{-20pt}
\psfrag{ks}{\raisebox{2ex}{\footnotesize\hspace{-0.8cm}$k_s$}}
\psfrag{a40}{\raisebox{0ex}{\footnotesize\hspace{0cm}$\ha_{4,0}$}}
\psfrag{g20}{\raisebox{0ex}{\footnotesize\hspace{0cm}$g_{2,0}$}}
  \includegraphics[width=3.0in]{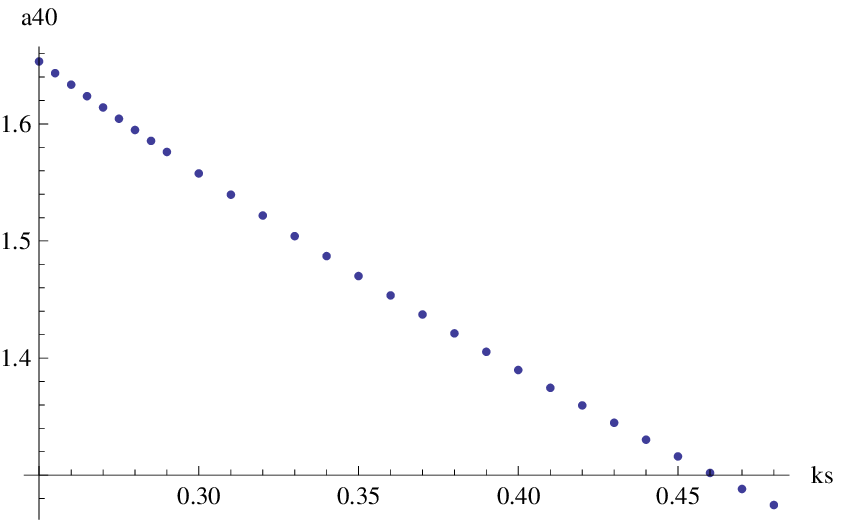}
  \includegraphics[width=3.0in]{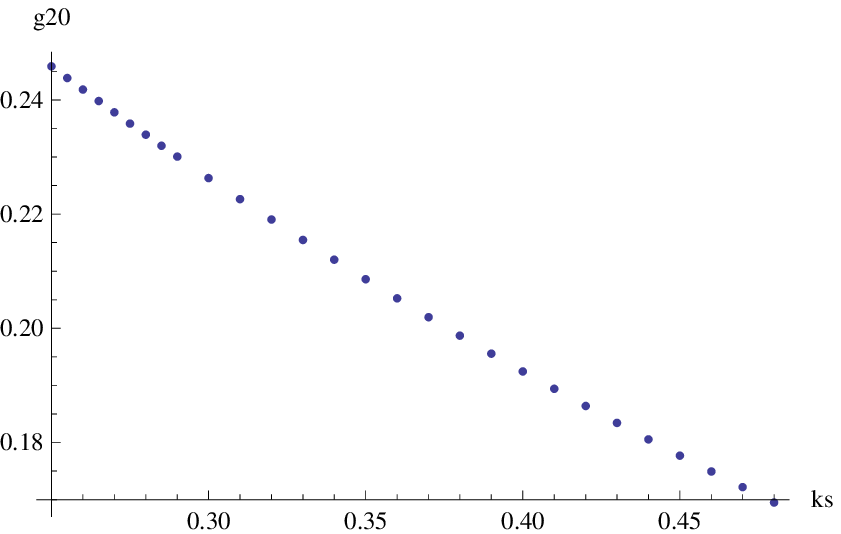}
  \caption{
The values of the UV parameters $\ha_{4,0}$ and $g_{2,0}$ as a
function of $k_s$.} \label{fig5}
\end{figure}

\begin{figure}[t]
 \hspace*{-20pt}
\psfrag{ks}{\raisebox{2ex}{\footnotesize\hspace{-0.8cm}$k_s$}}
\psfrag{a0h}{\raisebox{0ex}{\footnotesize\hspace{0cm}$a_0^h$}}
\psfrag{a1h}{\raisebox{0ex}{\footnotesize\hspace{0cm}$a_1^h$}}
  \includegraphics[width=3.0in]{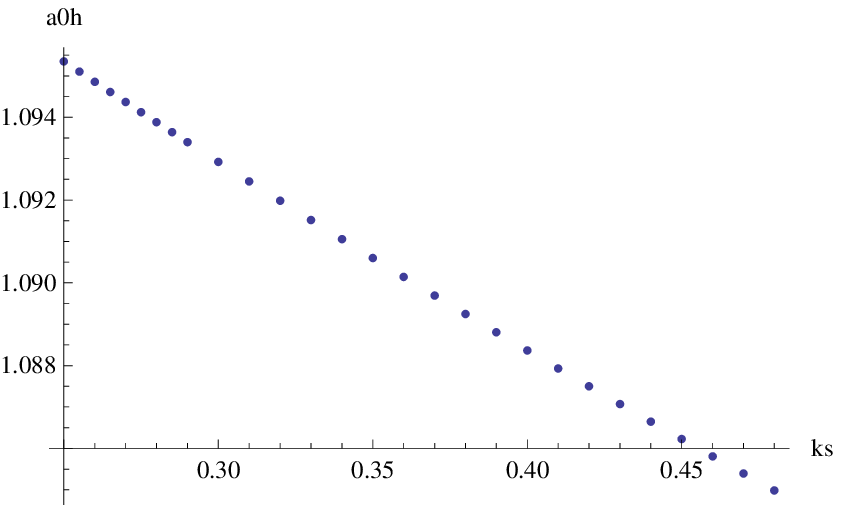}
  \includegraphics[width=3.0in]{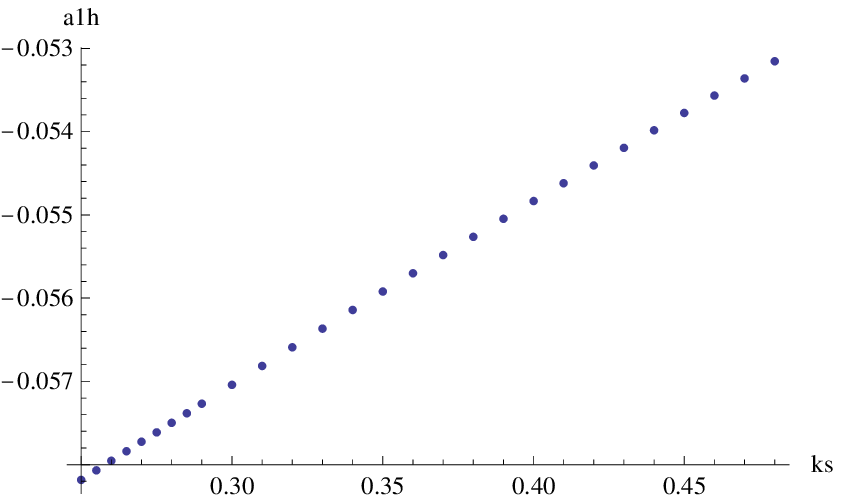}
  \caption{
The values of the IR parameters $a_0^h$ and $a_1^h$ as a function of
$k_s$.} \label{fig6}
\end{figure}

\begin{figure}[t]
 \hspace*{-20pt}
\psfrag{ks}{\raisebox{2ex}{\footnotesize\hspace{-0.8cm}$k_s$}}
\psfrag{b0h}{\raisebox{0ex}{\footnotesize\hspace{0cm}$b_0^h$}}
\psfrag{g0h}{\raisebox{0ex}{\footnotesize\hspace{0cm}$g_0^h$}}
  \includegraphics[width=3.0in]{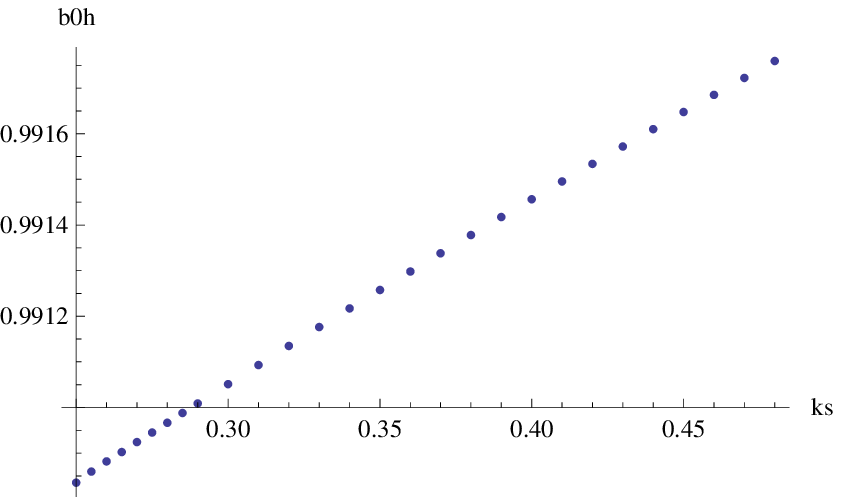}
  \includegraphics[width=3.0in]{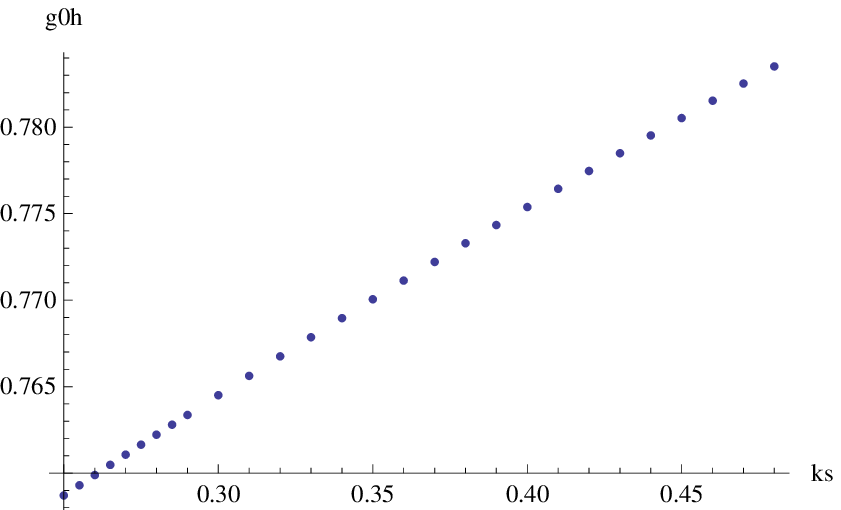}
  \caption{
The values of the IR parameters $b_0^h$ and $g_0^h$ as a function of
$k_s$.} \label{fig7}
\end{figure}

\begin{figure}[t]
 \hspace*{-20pt}
\psfrag{ks}{\raisebox{2ex}{\footnotesize\hspace{-0.8cm}$k_s$}}
\psfrag{k0h}{\raisebox{0ex}{\footnotesize\hspace{0cm}$k_0^h$}}
\psfrag{h0h}{\raisebox{0ex}{\footnotesize\hspace{0cm}$h_0^h$}}
  \includegraphics[width=3.0in]{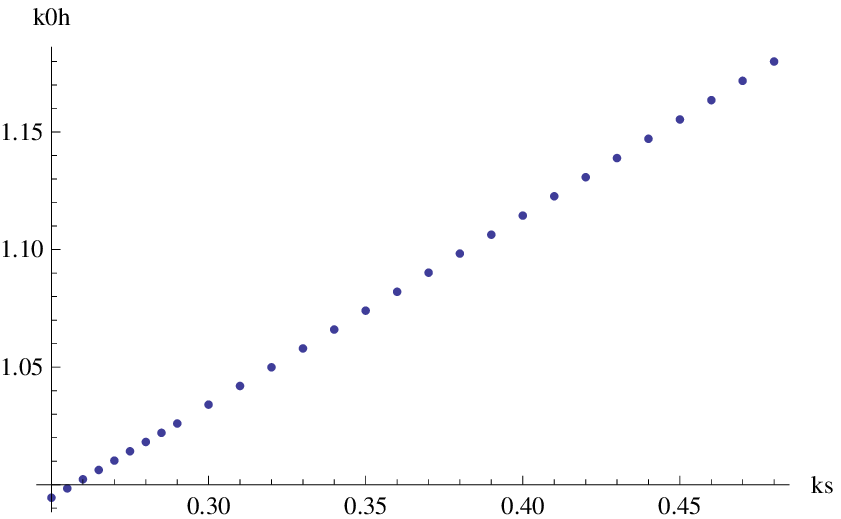}
  \includegraphics[width=3.0in]{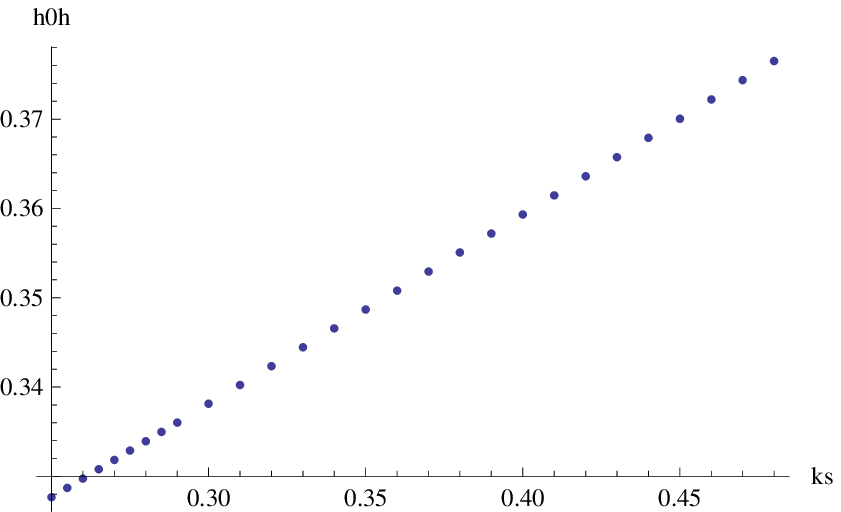}
  \caption{
The values of the IR parameters $k_0^h$ and $h_0^h$ as a function of
$k_s$.} \label{fig8}
\end{figure}

Figures \ref{fig4}-\ref{fig8} present the dependence of the UV
\{$\ha_{2,0}$, $\ha_{3,0}$, $\ha_{4,0}$, $g_{2,0}$\} and the IR
\{$a_0^h$, $a_1^h$, $b_0^h$, $g_0^h$, $k_0^h$, $h_0^h$\} parameters
on $k_s\in (0.25,0.48)$ with a step of $\Delta k_s=0.01$ ($\Delta
k_s=0.005$ near the transition) (blue points). In this regime the
typical norm of the mismatch vector \eqref{vmis}
$||\vec{v}_{mismatch}||\sim 10^{-5}$ or less. A highly non-trivial
check on our numerics is the consistency of the holographic flow
invariant $sT$. The latter can be computed in the IR using
\eqref{sT}, or in the UV using \eqref{bthermo}. We find that
\begin{equation}
\bigg|\frac{sT|_{IR}}{s T|_{UV}}-1\bigg|\sim 10^{-5}
\end{equation}
or less, which provides an independent check on the accuracy of matching the IR and UV solutions \eqref{vmis}.

The special (red) point in figure \ref{fig4} denotes a critical
value of $k_s=k_{critical}$, for which the corresponding value of
the parameter $\ha_{2,0}=\ha_{2,0}(k_{critical})=\frac {7}{12}$
leads to a vanishing of the free energy density \eqref{fenergy}. We
find $k_{critical}$ by performing a linear fit of the first 5
numerical points:
\begin{equation}
k_{critical}=0.25712(1)\,.
\end{equation}
Our available numerical data shows that the free energy density
\eqref{fenergy}
\begin{equation}
\calf=\frac{3 a_0^2}{28 \pi
G_5}\left(\ha_{2,0}-\frac{7}{12}\right)
\eqlabel{transition}
\end{equation}
is negative when $k_s > k_{critical}$ and is positive when $k_s <
k_{critical}$, so we find that at temperatures lower than
$k_s=k_{critical}$ the black hole solutions are not
thermodynamically preferred over the gas of particles in the
background of \cite{ks}. Thus, if we assume that these are the only
two possible configurations, $k_{critical}$ gives a critical
temperature corresponding to a first order confinement transition,
with chiral symmetry breaking, in the gravitational dual to the
cascading gauge theory. Examination of the infrared parameters in
figures \ref{fig6}-\ref{fig8} shows that the geometry at this
transition is non-singular (as expected for a first order
transition), and can be made arbitrarily weakly curved for large
values of $P$, justifying the validity of the supergravity
approximation. This observation is the main result of our paper.

\section{The physical results obtained from our numerical solutions}

In this final section we translate the results of the previous
section into physical quantities in the theory as a function of the
temperature. We present all the results as a function of
$T/\Lambda$, where the temperature $T$ is given by \eqref{sT}, and
the scale $\Lambda$ enters through the temperature dependence of
$k_s$. Recall that the numerical results presented in section
\ref{numerics} were obtained when setting $P=g_0=1$ and $a_0=1$. It
is easy to restore the correct powers of $P$ using the scaling
symmetry \eqref{scaling2}, and to then put a factor of $g_0=\hg_0$
together with every factor of $P^2$. In order to relax the $a_0=1$
condition, all the dimensionful quantities must be computed in units
of (see \eqref{lambdadef})
\begin{equation}
\Lambda=e^{-k_s/2}\,. \eqlabel{ldef}
\end{equation}
In particular, from \eqref{sT} we have
\begin{equation}
\frac{T}{\Lambda}=\frac{e^{k_s/2}}{4\pi h_0^h b_0^h}\ \sqrt{\frac{2
(8 h_0^h (a_0^h)^2-g_0^h)}{a_0^h+2 a_1^h}}\,, \eqlabel{sTl}
\end{equation}
enabling us to translate the dependence on $k_s$ into a dependence
on the temperature.

Equations \eqref{epressure} and \eqref{fenergy} imply that the
simplest expressions arise for the free energy density and the
energy density divided by $sT$, which are given by
\begin{equation}
\frac{\calf}{sT}=\frac 37\left(\ha_{2,0}-\frac{7}{12}\right)\,,
\qquad \frac{\cale}{sT}=\frac 34\left(1+\frac{4}{7}
\ha_{2,0}\right)\,. \eqlabel{efl}
\end{equation}
Equations \eqref{lambdadef} and \eqref{sT} allow us to compute the
entropy density divided by the temperature cubed, which is a measure
of the number of degrees of freedom in the theory :
\begin{equation}
\frac{4 \pi G_5}{P^4\hg_0^2} \frac{s}{T^3}=\frac{32\pi^4 \ sT}{81
M^4\hg_0^2\ T^4}=\frac{32\pi^4 \  sT}{81 M^4\hg_0^2\ \Lambda^4}
\left(\frac{\Lambda}{T}\right)^4= \left(\frac{1}{4\pi h_0^h
b_0^h}\ \sqrt{\frac{2 (8 h_0^h (a_0^h)^2-g_0^h)}{a_0^h+2
a_1^h}}\right)^{-4}\,. \eqlabel{stl}
\end{equation}
Notice that at high temperatures we can use the perturbative expression \eqref{a0} of Appendix \ref{hight} and \eqref{bthermo} to determine
\begin{equation}
\frac{s}{T^3}=\frac{sT}{T^4}=\frac{a_0^2}{4 \pi G_5
T^4}=\frac{\pi^4 K_\star^2}{64 \pi
G_5}\left(1+\calo\left(\frac{P^2\hg_0}{K_\star}\right)\right)
\simeq \frac{81}{128} M^4\hg_0^2\ \ln^2(\frac{T}{\Lambda})\,.
\eqlabel{hightst}
\end{equation}
Finally, we can evaluate the vacuum expectation values of the two
dimension 4 scalar operators \eqref{kpvev} :
\begin{equation}
\begin{split}
\frac{\langle\calo_{K_0}\rangle}{\Lambda^4}=&\frac{24}{7}\ \frac{e^{2k_s}}{P^2 \hg_0} \ \ha_{2,0}\,, \\
\frac{\langle\calo_{p_0}\rangle}{\Lambda^4}=&e^{2k_s}\left(2\
\frac{g_{2,0}}{\hg_0^2}+\frac{12(1-\ln(2) + 2 k_s)}{7 \hg_0}\
\ha_{2,0}\right)\,.
\end{split}
\eqlabel{kpvev1}
\end{equation}

\begin{figure}[t]
 \hspace*{-20pt}
\psfrag{lnTL}{\raisebox{2ex}{\footnotesize\hspace{-0.8cm}$\ln
(\frac{T}{\Lambda})$}}
\psfrag{ks}{\raisebox{0ex}{\footnotesize\hspace{0cm}$k_s$}}
  \includegraphics[width=3.0in]{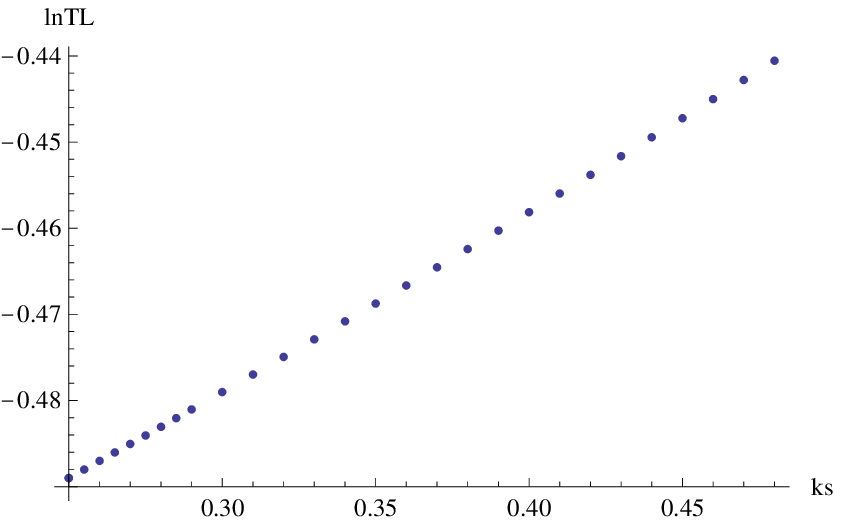}
  \includegraphics[width=3.0in]{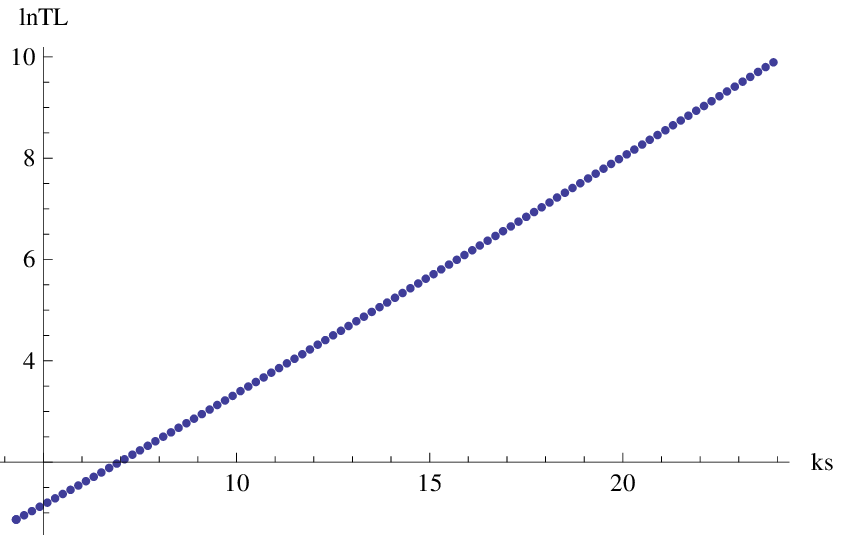}
  \caption{
The relation between $k_s$ and the temperature $T$. } \label{fig8a}
\end{figure}

\begin{figure}[t]
 \hspace*{-20pt}
\psfrag{TL}{\raisebox{2ex}{\footnotesize\hspace{-0.8cm}$\frac{T}{\Lambda}$}}
\psfrag{FsT}{\raisebox{0ex}{\footnotesize\hspace{0cm}$\frac{\calf}{sT}$}}
\psfrag{EsT}{\raisebox{0ex}{\footnotesize\hspace{0cm}$\frac{\cale}{sT}$}}
  \includegraphics[width=3.0in]{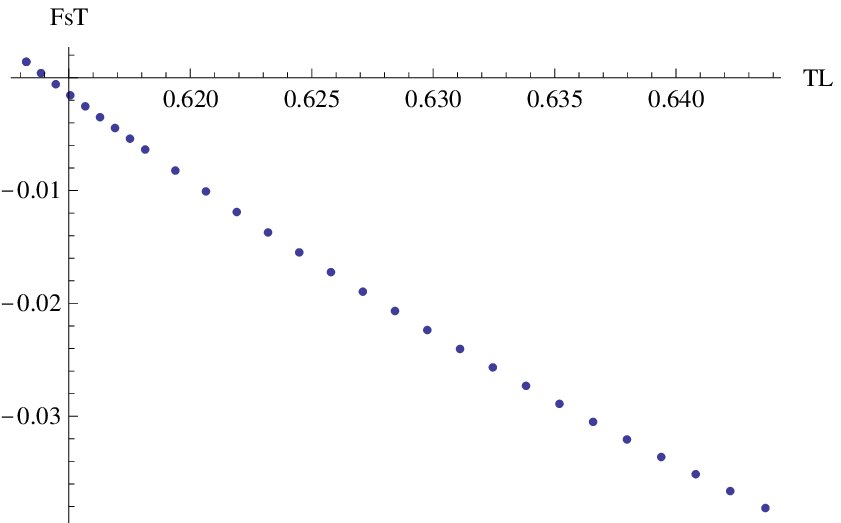}
  \includegraphics[width=3.0in]{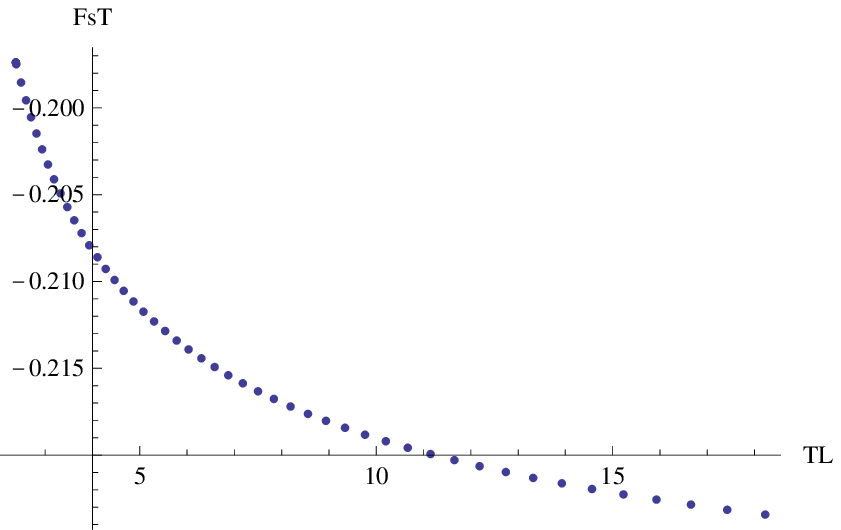}
  \caption{
The free energy density $\calf$, divided by $s T$, as a function of
$\frac{T}{\Lambda}$. On the left we plot temperatures at and
slightly above the deconfinement transition, and on the right much
higher temperatures.} \label{fig9}
\end{figure}

\begin{figure}[t]
 \hspace*{-20pt}
\psfrag{TL}{\raisebox{2ex}{\footnotesize\hspace{-0.8cm}$\frac{T}{\Lambda}$}}
\psfrag{FsT}{\raisebox{0ex}{\footnotesize\hspace{0cm}$\frac{\calf}{sT}$}}
\psfrag{EsT}{\raisebox{0ex}{\footnotesize\hspace{0cm}$\frac{\cale}{sT}$}}
  \includegraphics[width=3.0in]{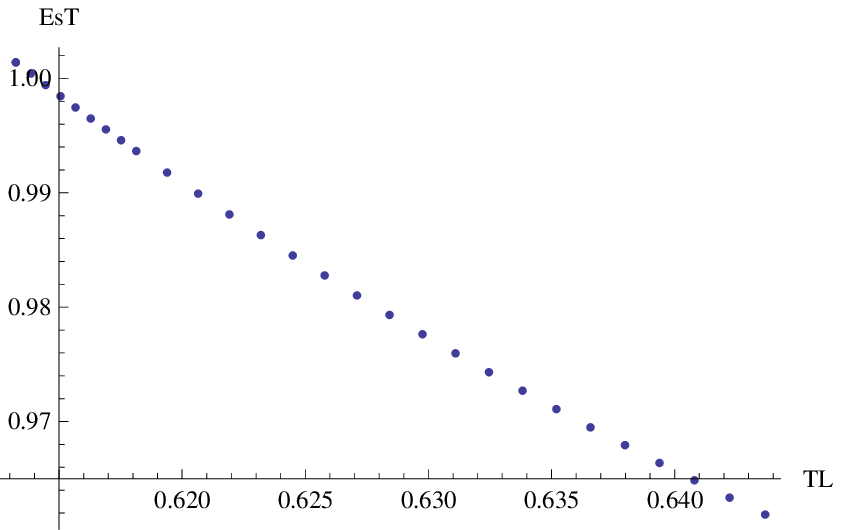}
  \includegraphics[width=3.0in]{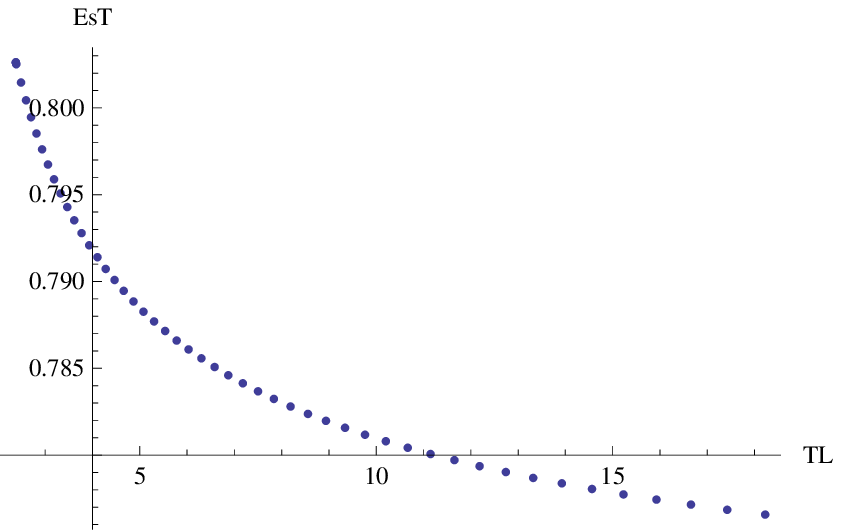}
  \caption{
The  energy density $\cale$, divided by $s T$, as a function of
$\frac{T}{\Lambda}$.} \label{fig9a}
\end{figure}

\begin{figure}[t]
 \hspace*{-20pt}
\psfrag{lnTL}{\raisebox{2ex}{\footnotesize\hspace{-0.8cm}$\ln\left(\frac{T}{\Lambda}\right)$}}
\psfrag{sT3half}{\raisebox{0ex}{\footnotesize\hspace{0cm}$\left(\frac{32\pi^4
\ s}{81 M^4\hg_0^2\ T^3}\right)^{1/2}\propto N_{eff} / M^{2}\hg_0$}}
  \includegraphics[width=3.0in]{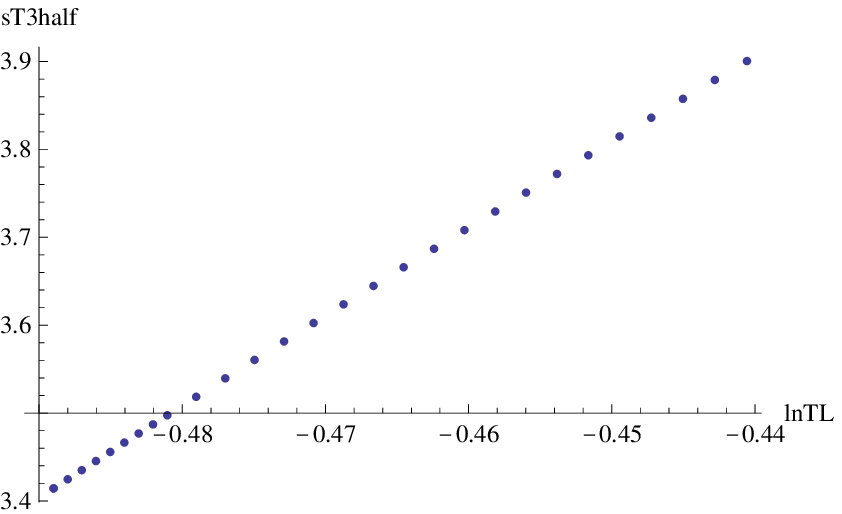}
  \includegraphics[width=3.0in]{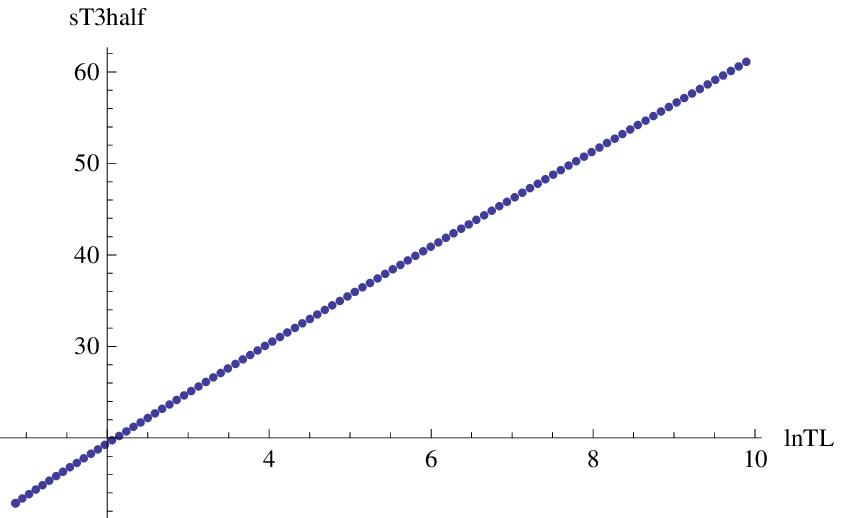}
  \caption{
The temperature dependence of the effective number of degrees of
freedom in the strongly coupled cascading gauge theory, as defined
by $N_{eff}^2\propto s/T^3$.} \label{fig10}
\end{figure}

\begin{figure}[t]
 \hspace*{-20pt}
\psfrag{TL}{\raisebox{2ex}{\footnotesize\hspace{-0.8cm}$\frac{T}{\Lambda}$}}
\psfrag{OK}{\raisebox{0ex}{\footnotesize\hspace{0cm}$P^{-2} \hg_0^{-1}\ \frac{\langle\calo_{K_0}\rangle}{\Lambda^4}$}}
\psfrag{Op}{\raisebox{0ex}{\footnotesize\hspace{0cm}$P^{-4} \hg_0^{-1}\ \frac{\langle\calo_{p_0}\rangle}{\Lambda^4}$}}
  \includegraphics[width=3.0in]{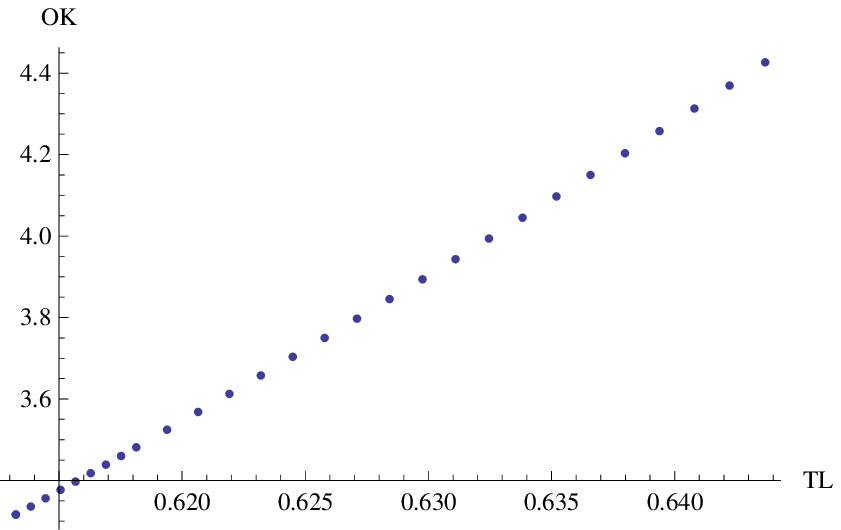}
  \includegraphics[width=3.0in]{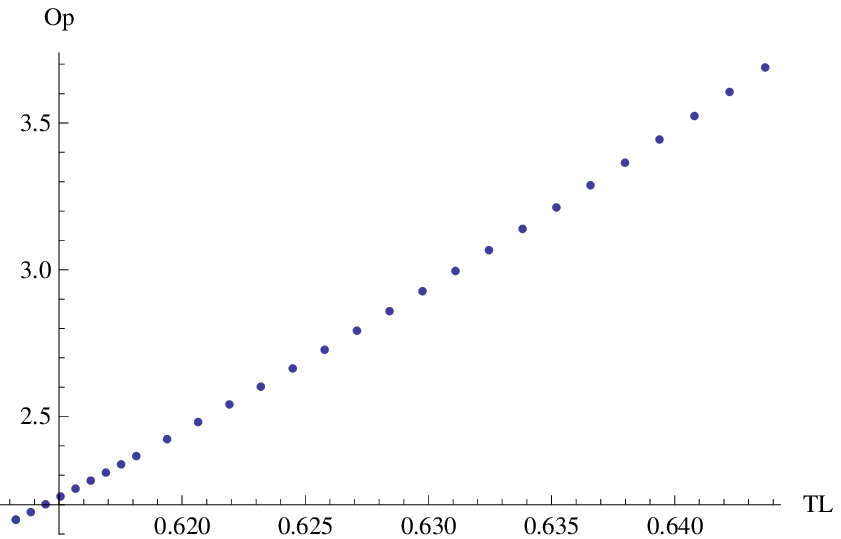}
  \caption{
The temperature dependence of the vacuum expectation values of the
dimension 4 operators $\langle\calo_{K_0}\rangle$  and
$\langle\calo_{p_0}\rangle$. The operators are normalized such that
they are invariant under the scaling transformation
\eqref{scaling2}. }\label{fig11}
\end{figure}

Figure \ref{fig8a} presents $\ln (\frac{T}{\Lambda})$ as a function
of $k_s$ at low and high temperatures. This is useful to determine
the temperature dependence of the various UV and IR parameters
presented in figures \ref{fig2}-\ref{fig8}. Notice that the high
temperature dependence of $k_s$ is in a good  agreement with the
high temperature asymptotic analysis of appendix \ref{hight}.
Indeed, a straight line fit of the points in the plot on the right
determines the slope to be $0.46(3)$, while the $k_s\to \infty$
slope is expected to be $\frac 12$ \eqref{leadks}.

Figure \ref{fig9} presents $\frac{\calf}{sT}$ as a function of
$\frac{T}{\Lambda}$ at low and high temperatures. Using a straight
line fit of the first 6 points in the (left) free energy density
plot, we determine the deconfinement and chiral symmetry restoration
temperature  to be
\begin{equation}
\left(\frac{T}{\Lambda}\right)_{critical}=0.614111(3)\,,
\eqlabel{tcrit}
\end{equation}
by requiring that the free energy density vanishes at
$T=T_{critical}$. Notice that there are noticeably large deviations
from scale invariant thermodynamics even for rather large
temperatures. Indeed, for $\frac{T}{\Lambda}\sim 10$, the deviation
of $\frac{\calf}{sT}$ from the conformal result
\begin{equation}
\frac{\calf}{sT}\bigg|_{conformal}=-\frac 14
\eqlabel{fstc}
\end{equation}
is about $12\%$.

Figure \ref{fig9a} presents $\frac{\cale}{sT}$  as a function of
$\frac{T}{\Lambda}$ at low and high temperatures. Here, the
deviation at high temperature from the conformal result
\begin{equation}
\frac{\cale}{sT}\bigg|_{conformal}=\frac 34
\eqlabel{estc}
\end{equation}
is three times less than the corresponding deviation in the free
energy density (or pressure). Such a suppression is easy to
understand once we notice from \eqref{efl} that
\begin{equation}
\frac{\calf}{sT}=\biggl(1-\delta\biggr)\times
\frac{\calf}{sT}\bigg|_{conformal}\,, \qquad
\frac{\cale}{sT}=\left(1+\frac 13 \delta\right)\times
\frac{\cale}{sT}\bigg|_{conformal}\,, \eqlabel{deviate}
\end{equation}
where $\delta\equiv \frac{12}{7} \ha_{2,0}$. Note that the lattice
results for QCD also imply that the energy density of the QCD plasma
near the deconfinement transition is much more similar to that of
scale-invariant thermodynamics than the QCD pressure \cite{lattice}.

Figure \ref{fig10} presents the temperature dependence of the
effective number of degrees of freedom of the strongly coupled
cascading gauge theory at low and high temperatures, as defined by
$N_{eff}^2\propto s/T^3$. One way to characterize the phase
transition temperature is by the effective number of degrees of
freedom (in the deconfined phase) at this temperature. Using the
straight line fit of the first $6$ points in the (left) effective
number of degrees of freedom plot, we find that at the deconfinement
and chiral symmetry restoration temperature \eqref{tcrit}
\begin{equation}
\frac{32\pi^4}{81 M^4\hg_0^2}
\frac{s}{T^3}\bigg|_{T=T_{critical}}=3.4291(5). \eqlabel{neffcrit}
\end{equation}

Figure \ref{fig11} presents the temperature dependence of the vacuum
expectation values of the dimension 4 operators
$\langle\calo_{K_0}\rangle$ and $\langle\calo_{p_0}\rangle$ (see
\eqref{kpvev1}) at low temperatures. At high temperatures we can use
the perturbative expressions \eqref{a0}, \eqref{a20a0} and
\eqref{g20}  of appendix \ref{hight}  to determine
\begin{equation}
\begin{split}
P^{-2} \hg_0^{-1}\ \frac{\langle\calo_{K_0}\rangle}{\Lambda^4}&\propto  \left(\frac{T}{\Lambda}\right)^4
\ln (\frac{T}{\Lambda})\,,\\
P^{-4} \hg_0^{-1}\
\frac{\langle\calo_{p_0}\rangle}{\Lambda^4}&\propto
\left(\frac{T}{\Lambda}\right)^4 \ln^2 (\frac{T}{\Lambda})\,.
\end{split}
\eqlabel{kpvev2}
\end{equation}

\section*{Acknowledgments}

We would like to thank Matthew Headrick, Igor Klebanov, Jim Liu, Chris
Pagnutti, Leo Pando-Zayas, Larry Yaffe and Amos Yarom for interesting
discussions. OA would like to thank the Institute for Advanced Study
for hospitality during the course of this project.  The work of OA is
supported in part by the Israel-U.S. Binational Science Foundation, by
a center of excellence supported by the Israel Science Foundation
(grant number 1468/06), by the European network HPRN-CT-2000-00122, by
a grant from the G.I.F., the German-Israeli Foundation for Scientific
Research and Development, and by a grant of DIP (H.52). AB would like
to thank the University of Texas at Austin for the hospitality, where
part of this work was done. AB's research at Perimeter Institute is
supported in part by the Government of Canada through NSERC and by the
Province of Ontario through MEDT. AB gratefully acknowledges further
support by an NSERC Discovery grant.  PK would like to thank the
Landesstiftung Baden-Wuerttemberg for the scholarship.

\appendix

\section{Appendix : Perturbative evaluation of the solutions at high
temperature}
\label{hight}

In this appendix we analyze the  high temperature thermodynamics of
the cascading gauge theory perturbatively in $P^2 \hg_0 / K_\star$,
where
\begin{equation}
K_\star=P^2\hg_0 \left(k_s +\frac 12 \ln
(2)+\calo\left(k_s^{-1}\right)\right) \eqlabel{ksdef}
\end{equation}
is the five-form flux evaluated at the horizon, and $k_s$ is defined by \eqref{num2}.

The purpose of this perturbative analysis is twofold: first, we
would like to test our asymptotic identification of the cascading
geometry parameters \eqref{match1}, of the temperature and of the
dynamical scale $\Lambda$, against the first law of
thermodynamics; second, we would like to obtain analytic
predictions for
the high-temperature values of the UV \eqref{uvpar} and the IR
parameters \eqref{irrap} perturbatively in $P^2\hg_0/K_\star$, in
order to benchmark our general numerical analysis. We will test the
first law of thermodynamics to order
$\calo\left(\frac{P^6\hg_0^3}{K_\star^3}\right)$ inclusive, and
evaluate the UV parameters \eqref{uvpar} to order
$\calo\left(\frac{P^4\hg_0^2}{K_\star^2}\right)$ inclusive.

The appendix is organized as follows. In subsection \ref{aequation}
we derive the perturbative equations of motion to order
$\calo\left(\frac{P^6\hg_0^3}{K_\star^3}\right)$ inclusive. In
subsection \ref{aexpansion} we present the near horizon and the near
boundary expansions of the solutions, outline our numerical method
for the computation of the UV/IR parameters of the perturbative
solutions, and collect numerical expressions for some of these
parameters. In subsection \ref{aphysics} we present perturbative
expressions for the thermodynamics of the deconfined cascading gauge
theory, and verify the first law of thermodynamics.

\subsection{Perturbative equations of motion}
\label{aequation}

As discussed above, without loss of generality we can set
$g_0=\hg_0=1$. We use the following parametrization for the
solution in perturbation theory in $\frac{P^2}{K_\star}$ :
\begin{equation}
\begin{split}
&h(x)= \frac {K_{\star}}{4\tilde{a}_0^2}+\frac{K_\star}{\ta_0^2}\  \sum_{n=1}^3 \left\{\left(\frac{P^2}{K_\star}\right)^n
\left(\xi_{2n}(x)-\frac 54 \eta_{2n}(x)\right)\right\}+\frac{K_\star}{\ta_0^2}\  \calo\left(\frac{P^8}{K_\star^4}\right)\,,\\
\end{split}
\eqlabel{p2order1}
\end{equation}
\begin{equation}
\begin{split}
&f_2(x)=\tilde{a}_0+\ta_0\ \sum_{n=1}^3 \left\{\left(\frac{P^2}{K_\star}\right)^n  \left(-2 \xi_{2n}(x)+\eta_{2n}(x)+\frac 45 \lambda_{2n}(x)\right)
\right\}+\ta_0\ \calo\left(\frac{P^8}{K_\star^4}\right)\,,\\
\end{split}
\eqlabel{p2order2}
\end{equation}
\begin{equation}
\begin{split}
&f_3(x)=\tilde{a}_0+\ta_0\ \sum_{n=1}^3 \left\{\left(\frac{P^2}{K_\star}\right)^n  \left(-2 \xi_{2n}(x)+\eta_{2n}(x)-\frac 15 \lambda_{2n}(x)\right)
\right\}+\ta_0\ \calo\left(\frac{P^8}{K_\star^4}\right)\,,\\
\end{split}
\eqlabel{p2order3}
\end{equation}
\begin{equation}
\begin{split}
&K(x)=K_{\star}+K_{\star}\ \sum_{n=1}^3 \left\{\left(\frac{P^2}{K_\star}\right)^n \k_{2n}(x)\right\}+K_\star\ \calo\left(\frac{P^8}{K_\star^4}\right)\,,\\
\end{split}
\eqlabel{p2order4}
\end{equation}
\begin{equation}
\begin{split}
&g(x)=1+\sum_{n=1}^3 \left\{\left(\frac{P^2}{K_\star}\right)^n \zeta_{2n}(x)\right\}+\calo\left(\frac{P^8}{K_\star^4}\right)\,.\\
\end{split}
\eqlabel{p2order5}
\end{equation}

The advantage of this parametrization is that the equations for
$\{\xi_{2n},\eta_{2n},\lambda_{2n},\zeta_{2n}\}$ decouple, once the
(decoupled) equation for $\k_{2n}$ is solved, at each order
($n=1,2,3$) in perturbation theory. We find (for $n=1,2,3$)
\begin{equation}
0=\k_{2n}''+\frac{\k_{2n}'}{x-1}+\calj_\k^{[2n]}\,, \eqlabel{eqk2n}
\end{equation}
\begin{equation}
0=\eta_{2n}''+\frac{\eta_{2n}'}{x-1}-\frac{8\eta_{2n}}{x^2(x-2)^2}-\frac
25 \k_2'\ \k_{2n}'-\frac{8\k_{2n}}{3x^2(x-2)^2} +\calj_\eta^{[2n]}\,,
\eqlabel{eqeta2n}
\end{equation}
\begin{equation}
0=\xi_{2n}''+\frac{(3x^2-6x+4)\xi_{2n}'}{x(x-1)(x-2)}-\frac 23
\k_2'\ \k_{2n}'+\calj_\xi^{[2n]}\,, \eqlabel{eqxi2n}
\end{equation}
\begin{equation}
0=\lambda_{2n}''+\frac{\lambda_{2n}'}{x-1}-\frac{3\lambda_{2n}}{x^2(x-2)^2}-2
\k_2'\ \k_{2n}'+\calj_\lambda^{[2n]}\,, \eqlabel{eqlambda2n}
\end{equation}
\begin{equation}
0=\zeta_{2n}''+\frac{\zeta_{2n}'}{x-1}+2 \k_2'\
\k_{2n}'+\calj_\zeta^{[2n]}\,, \eqlabel{eqzeta2n}
\end{equation}
where the source terms
$\{\calj_\k^{[2n]},\calj_\eta^{[2n]},\calj_\xi^{[2n]},\calj_\lambda^{[2n]},\calj_\zeta^{[2n]}\}$
are functionals of the lower order solutions: $\k_{2m}$, $\xi_{2m}$,
$\eta_{2m}$, $\lambda_{2m}$, $\zeta_{2m}$, with $m<n$. Explicit
expressions for the source term functionals are available from the
authors upon request.

The perturbative solutions to  \eqref{eqk2n}-\eqref{eqzeta2n} must
be regular at the horizon, and must have the appropriate KT
asymptotics \eqref{genexp} near the boundary.

The leading order ($n=1$) solution to \eqref{eqk2n}-\eqref{eqzeta2n}
was obtained in \cite{bh2}:
\begin{equation}
\k_2=-\frac 12 \ln (2x-x^2)\,, \eqlabel{ksol}
\end{equation}
\begin{equation}
\eta_2=\frac{(x^2-2x+2)}{20(2x-x^2)}\left(\dilog(2x-x^2)-\frac
16\pi^2\right)-\frac{1}{15}+\frac{\ln(2x-x^2)}{15}\,, \eqlabel{etas}
\end{equation}
\begin{equation}
\xi_2=\frac{1}{12}\ln(2x-x^2)\,, \eqlabel{xis}
\end{equation}
\begin{equation}
\zeta_2=\frac{\pi^2}{24}-\frac 12 \dilog(1-x)+\frac 12
\dilog(2-x)-\frac 12 \ln (x)\ \ln (1-x)\,. \eqlabel{zetas}
\end{equation}
There is no simple expression for $\l_2$ --- it is straightforward
to write an appropriate solution using the Green's function for
\eqref{eqlambda2n}, but this explicit expression is not useful.

Similarly, although the higher order $n=2,3$ solutions to
\eqref{eqk2n}-\eqref{eqzeta2n} could be presented in quadratures,
these expressions are not useful. Rather, we identify the higher
order solutions by specifying their asymptotic expressions near
the horizon and near the boundary, along with the numerical values
for the relevant integration constants.

\subsection{UV/IR asymptotics of the perturbative solutions}
\label{aexpansion}

The UV/IR parameters of the higher order perturbative solutions are
found by solving the differential equations
\eqref{eqk2n}-\eqref{eqzeta2n} numerically from the boundary
($x=0$), and requiring the proper boundary conditions at the
horizon, which are
\begin{equation}
\lim_{x\to _{1_-}}\k'_{2n}=\lim_{x\to _{1_-}}\eta'_{2n}=\lim_{x\to
_{1_-}}\xi'_{2n}=\lim_{x\to _{1_-}}\l'_{2n}=\lim_{x\to
_{1_-}}\zeta'_{2n}=0\,,\qquad n=2,3\,. \eqlabel{hbc}
\end{equation}

To begin, we present the asymptotics of $\l_2$. As $x\to 0_+$ we
find
\begin{equation}
\lambda_2=\frac 23 x+\l_3^{[2]} x^{3/2}+\frac{11}{15} x^2+\frac 34
\l_3^{[2]} x^{5/2}+\frac{176}{315} x^3+\frac{9}{16} \l_3^{[2]}
x^{7/2}+\frac{676}{1575} x^4+\frac{7}{16} \l_3^{[2]}
x^{9/2}+\calo\left(x^5\right)\,, \eqlabel{l0sol}
\end{equation}
where $\l_3^{[2]}$ is related to the condensate of the dimension 6
operator at order $\calo\left(\frac{P^2}{K_\star}\right)$. Nonsingularity of the $\l_2(x)$
solution to \eqref{eqlambda2n} at the horizon, together with
\[\lim_{x\to _{1_-}}\l'_{2}=0\,,\] determines
\begin{equation}
\l_3^{[2]}=-0.872358024(9)\,.
\end{equation}
Near the horizon, as $y\equiv 1-x\to 0_+$, we find
\begin{equation}
\begin{split}
\lambda_2=&\l_{2,0}^h+\left(-\frac 14+\frac 34 \l_{2,0}^h\right)
y^2+\left(\frac{33}{64} \l_{2,0}^h-\frac{7}{64}\right) y^4
+\left(\frac{107}{256} \l_{2,0}^h-\frac{181}{2304}\right) y^6\\
&+\left(-\frac{3181}{49152}+\frac{5913}{16384} \l_{2,0}^h\right)
y^8+\calo\left(y^{10}\right)\,,
\end{split}
\eqlabel{l1sol}
\end{equation}
where $\l_{2,0}^h$ can be determined numerically to be
\begin{equation}
\lambda_{2,0}^h=0.16806(9)\,.
\end{equation}

\subsubsection{Order $n=2$ asymptotics}
\
\nxt As $x\to 0_+$ we find
\begin{equation}
\begin{split}
\k_4=&\left(\k_{2}^{[4]}-\frac 12 \ln (x)\right) x
+\frac{2}{15}\l_3^{[2]}\ x^{3/2}
+\left(-\frac{106}{225}+\frac{7}{30} \ln (2)
+\frac 12 \k_{2}^{[4]}-\frac{1}{60} \ln (x)\right) x^2\\
&+\frac{1}{50}\l_3^{[2]}\ x^{5/2}+\calo\left(x^3\right)\,,
\end{split}
\eqlabel{k4h}
\end{equation}
\begin{equation}
\begin{split}
\eta_4=&\left(-\frac{1}{12}+\frac{1}{18} \ln (2)+\frac{1}{18} \ln
(x)\right) +\left(\frac{13}{360}-\frac{1}{30} \ln (2)-\frac{7}{30}
\k_{2}^{[4]}+\frac{1}{12} \ln (x)\right) x
\\
&-\frac{4}{225}\l_3^{[2]}\
x^{3/2}+\left(\eta_{4}^{[4]}+\left(\frac{7}{360}-\frac{1}{36} \ln
(2)+\frac{1}{15} \k_{2}^{[4]}\right) \ln (x)
-\frac{11}{360} \ln^2 (x)\right) x^2\\
&-\frac{97}{1575}\l_3^{[2]}\ x^{5/2}+\calo\left(x^3\right)\,,
\end{split}
\eqlabel{e4h}
\end{equation}
\begin{equation}
\begin{split}
&\xi_4=\frac{1}{36} \ln (x)+\left(-\frac 16
\k_{2}^{[4]}+\frac{11}{144}
-\frac{1}{24} \ln (2)+\frac{1}{24} \ln (x)\right) x-\frac{2}{225}\l_3^{[2]}\ x^{3/2}\\
&+\left(-\frac{191}{43200}-\frac{1}{36} \k_{2}^{[4]}+\frac{1}{80}
\ln (2)-\frac{1}{144} \ln^2 (2)
+\biggl(\frac{19}{720}-\frac{1}{72} \ln (2)\right) \ln (x)\\
&-\frac{1}{144} \ln^2 (x)\biggr) x^2
-\frac{229}{3150}\l_3^{[2]}\ x^{5/2}+\calo\left(x^3\right)\,,
\end{split}
\eqlabel{x4h}
\end{equation}
\begin{equation}
\begin{split}
\lambda_4=&\left(-\frac{14}{9}+\frac 43 \k_{2}^{[4]}+\frac 23 \ln
(2)\right) x+\l_3^{[4]}\ x^{3/2}
+\left(\frac{67}{450}+\frac{22}{15} \k_{2}^{[4]}
+\frac 13 \ln (2)-\frac 25 \ln x\right) x^2\\
&+\left(\frac 45 \l_3^{[2]}+\frac 34 \l_3^{[4]}\right)
x^{5/2}+\calo\left(x^3\right)\,,
\end{split}
\eqlabel{l4h}
\end{equation}
\begin{equation}
\begin{split}
\zeta_4=&\left(\zeta_{2}^{[4]}+\left(-\frac{13}{12}+\k_{2}^{[4]}+\frac 12 \ln (2)\right) \ln (x)\right) x\\
&+\left( -\frac{13}{24}+\frac 12 \zeta_{2}^{[4]}+\frac 18 \ln^2
(2)+\left(-\frac{13}{24}+\frac 12 \k_{2}^{[4]}+\frac 12 \ln
(2)\right) \ln (x)
+\frac 18 \ln^2 (x)\right) x^2\\
&-\frac{4}{75} \l_3^{[2]}\ x^{5/2} +\calo\left(x^3\right)\,,
\end{split}
\eqlabel{z4h}
\end{equation}
where the new UV parameters
\{$\k_{2}^{[4]},\eta_{4}^{[4]},\l_{3}^{[4]},\zeta_{2}^{[4]}$\}
are determined numerically from the horizon boundary condition \eqref{hbc}:
\begin{equation}
\k_{2}^{[4]}\bigg|_{numeric}=0.73675974(3)\,,
\eqlabel{k20n}
\end{equation}
and
\begin{equation}
\begin{split}
&\eta_{4}^{[4]}\bigg|_{numeric}=0.0053421556(6)\,,\\
&\l_{3}^{[4]}\bigg|_{numeric}=-1.1156300100(2)\,,\\
&\z_{2}^{[4]}\bigg|_{numeric}=0.622262593(4)\,.
\end{split}
\eqlabel{othern}
\end{equation}
With \eqref{k20n} and \eqref{othern} there are no additional UV
parameters  to tune in order to enforce the horizon boundary
condition \eqref{hbc} for $\xi_{4}$. We find
\begin{equation}
\xi'_{4}(x=0.99999)\bigg|_{numeric}\sim 10^{-6}\,, \eqlabel{xih1}
\end{equation}
which is of the same order of magnitude as the error in \eqref{hbc}
for all other functions. \nxt The asymptotic  expressions of the
$n=2$ solutions near the horizon $y\equiv 1-x\to 0_+$ take the form
\begin{equation}
\begin{split}
\k_4=&\k_{4,0}^h+\k_{4,2}^h y^2+\k_{4,4}^h y^4+\calo(y^6)\,,\\
\eta_4=&\eta_{4,0}^h+\eta_{4,2}^h y^2+\eta_{4,4}^h y^4+\calo(y^6)\,,\\
\xi_4=&\xi_{4,0}^h+\xi_{4,2}^h y^2+\xi_{4,4}^h y^4+\calo(y^6)\,,\\
\l_4=&\l_{4,0}^h+\l_{4,2}^h y^2+\l_{4,4}^h y^4+\calo(y^6)\,,\\
\zeta_4=&\zeta_{4,0}^h+\zeta_{4,2}^h y^2+\zeta_{4,4}^h
y^4+\calo(y^6)\,.
\end{split}
\eqlabel{forth0}
\end{equation}
To verify the first law of thermodynamics to order
$\calo\left(\frac{P^6}{K\star^3}\right)$ we will need the
numerical expressions only for
$\{\k_{4,0}^h,\xi_{4,0}^h,\xi_{4,2}^h\}$. We find
\begin{equation}
\begin{split}
&\k_{4,0}^h\bigg|_{numeric}=0.62226(3)\,,
\end{split}
\eqlabel{hor1}
\end{equation}
and
\begin{equation}
\begin{split}
&\xi_{4,0}^h\bigg|_{numeric}=-0.079819(3)\,,\\
&\xi_{4,2}^h\bigg|_{numeric}=0.019198(8)\,.\\
\end{split}
\eqlabel{hor}
\end{equation}

\subsubsection{Order $n=3$ asymptotics}

To verify the first law of thermodynamics to order
$\calo\left(\frac{P^6}{K\star^3}\right)$ we will need the
asymptotic expression for $\k_6$ only. We find
\begin{equation}
\begin{split}
\k_6=& \left(\k_2^{[6]}+\left(\frac{13}{12}-\k_2^{[4]}-\frac 12
\ln (2)\right) \ln (x)\right)x
+\left(\frac{1}{45} \l_3^{[2]}+\frac{2}{15} \l_3^{[4]}\right)x^{3/2}\\
&+ \biggl(\frac{281821}{216000} -\frac{2587}{3600}
\k_2^{[4]}-\frac{509}{720} \ln (2)+\frac{31}{288} \ln^2 (2)+\frac
12 \k_2^{[6]}+\frac{13}{4}
\eta_4^{[4]}+\frac14 \z_2^{[4]}\\
&+\frac 14 \ln (2)\ \k_2^{[4]}
+\left(-\frac{49}{3600}+\frac{13}{60} \k_2^{[4]}
-\frac 18 \ln (2)\right) \ln (x)-\frac{7}{60} \ln^2 (x)\biggr)x^2\\
&+ \left(\frac{1153}{10500} \l_3^{[2]}-\frac{4}{25} \l_3^{[2]}
\k_2^{[4]}+\frac{1}{50} \l_3^{[4]}+\frac{2}{25} \l_3^{[2]} \ln
(x)\right)x^{5/2} +\calo\left(x^3\right)
\end{split}
\eqlabel{k6h}
\end{equation}
as $x\to 0_+$, with
\begin{equation}
\k_2^{[6]}=-0.62226259(3)
\eqlabel{k26}
\end{equation}
determined from the horizon boundary condition \eqref{hbc}.

\subsubsection{Perturbative expressions for UV parameters \eqref{genexp}}

Finally, we collect perturbative expressions for the various
independent UV parameters  $a_{2,0}$, $a_{3,0}$, $a_{4,0}$,
$g_{2,0}$ as defined by \eqref{genexp}. Because of the scaling
symmetry \eqref{scaling} it is convenient to quote these parameters
relative to $a_0$. Also, in the next subsection we show that the
first law of thermodynamics requires that
\begin{equation}
\k_{2}^{[4]}=\frac{13}{12}-\frac 12\ln(2)\,. \eqlabel{k10p}
\end{equation}
Note that \eqref{k10p} agrees with \eqref{k20n} up to an error of
order $10^{-10}$. In the following expressions for the UV parameters
in \eqref{genexp} we use the analytic expression \eqref{k10p}.

We find
\begin{equation}
\begin{split}
\frac{a_{2,0}}{a_0}=\frac{7}{12}\
\frac{P^2}{K_\star}+\calo\left(\frac{P^6}{K_\star^3}\right)\,,
\end{split}
\eqlabel{a20a0}
\end{equation}
\begin{equation}
\begin{split}
\frac{a_{3,0}}{a_0}=&\frac 45 \l_3^{[2]}\
\frac{P^2}{K_\star}+\left(\frac 45
\l_3^{[4]}+\frac{2}{15}\l_3^{[2]}\right)\ \frac{P^4}{K_\star^2}
+\calo\left(\frac{P^6}{K_\star^3}\right)\,,
\end{split}
\eqlabel{a30a0}
\end{equation}
\begin{equation}
\begin{split}
\frac{a_{4,0}}{a_0}=&\left(\frac{1}{30}\ln
(2)+\frac{1021}{1800}\right)\ \frac{P^2}{K_\star}
+\left(\eta_4^{[4]}-\frac{661}{1800}\ln (2)+\frac {1}{72}
\left(\ln (2)\right)^2+\frac{167809}{108000}\right)\
\frac{P^4}{K_\star^2}
\\
&+\calo\left(\frac{P^6}{K_\star^3}\right)\,,
\end{split}
\eqlabel{a40a0}
\end{equation}
\begin{equation}
\begin{split}
g_{2,0}=&\left(-\frac 12+\frac 12 \ln (2)\right)\
\frac{P^2}{K_\star}+\z_2^{[4]}\ \frac{P^4}{K_\star^2}
+\calo\left(\frac{P^6}{K_\star^3}\right)\,.
\end{split}
\eqlabel{g20}
\end{equation}

\subsection{Perturbative thermodynamics of the non-extremal cascading geometry}
\label{aphysics}

One of the interesting properties of the deconfined cascading
geometry is the temperature dependence of the five-form flux
evaluated at the horizon $K_\star=K_{\star}(T)$. The need for such
dependence was first pointed out in \cite{bh2}; it stems from the
fact that when studying the thermodynamics of non-conformal gauge
theories (such as the cascading gauge theory) one must
keep the intrinsic scale of the cascading gauge theory
fixed\footnote{This fact was not clearly taken into account in
previous numerical studies \cite{lpz}.
}, rather than keeping fixed the five-form at the horizon.

The fact that $K_\star$ is temperature dependent introduces
additional temperature dependence into the thermodynamic
potentials (the free energy density $\calf$ \eqref{fenergy}, the
energy density $\cale$ \eqref{epressure}, and the entropy density
$s$ \eqref{sT}) via the UV parameters $a_0$ and $a_{2,0}$, both of
which depend on $K_\star$.
As a result, the first law of thermodynamics
\begin{equation}
d\calf=-s\ dT
\eqlabel{first1}
\end{equation}
would not be valid, unless the temperature dependence of $K_\star$
is properly determined and taken into account. One possible approach
is to use the first law of thermodynamics \eqref{first1} as a way to
determine $K_\star(T)$. Such an approach was proposed and
implemented in \cite{aby} to leading order in
$\calo\left(\frac{P^2}{K_\star}\right)$, where it was found that
validity of \eqref{first1} requires that
\begin{equation}
\frac{dK_\star(T)}{dT}=\frac{2P^2}{T}+\calo\left(\frac{P^4}{K_\star}\right)
\eqlabel{leadks}
\end{equation}
(this was also proposed in \cite{bh2}, based on the requirement of
keeping the glueball spectrum scale of the cascading gauge theory
fixed). Equation \eqref{leadks} was also shown to be required for
the consistency of the hydrodynamics of the cascading gauge theory
plasma in \cite{tranc}.

The main observation of this paper is that one can rigorously
determine the temperature dependence of $K_\star$ without referring
to the first law of thermodynamics. In the bulk of the paper this
was implicitly done in our solutions. In the context of the
perturbative high-temperature expansion, we can obtain such an
identification perturbatively by expanding the exact matching
condition \eqref{match1}, enforcing the fixed scale of the cascading
gauge theory, perturbatively in $\frac{P^2}{K_\star}$. We will
demonstrate here that this identification is consistent with the
first law of thermodynamics. This provides a non-trivial consistency
check on our solutions.

In the rest of this subsection we present explicit expressions for
$a_0$ as a function of the temperature $T$ to order
$\calo\left(\frac{P^4}{K_\star^2}\right)$. One can then use
\eqref{a20a0} to compute the thermodynamic potentials of the
cascading black hole geometry. We present explicit perturbative
expressions for $\frac{dK_\star(T)}{dT}$, and verify the first law
of thermodynamics to order
$\calo\left(\frac{P^6}{K_\star^3}\right)$.

\subsubsection{Cascading black hole thermodynamics to order $\calo\left(\frac{P^2}{K_\star}\right)$}

Explicitly evaluating the temperature of the black hole to order
$\calo\left(\frac{P^2}{K_\star}\right)$ we find
\begin{equation}
\tilde{a}_0=\frac{\pi^2K_\star T^2}{4}\
\left(1+\frac{2P^2}{3K_\star}+\calo\left(\frac{P^4}{K_\star^2}\right)\right)\,.
\eqlabel{ta0}
\end{equation}
Using \eqref{ksol}-\eqref{zetas}, we further determine
\begin{equation}
a_0=\frac{\pi^2K_\star T^2}{4}\ \left(1+\frac
{P^2}{2K_{\star}}+\calo\left(\frac{P^4}{K_\star^2}\right)\right)\,.
\eqlabel{a0}
\end{equation}
The matching condition \eqref{match1} then determines
\begin{equation}
\calo\left(\frac{P^4}{K_\star}\right)
=4\hh_{0,0}-K_\star+P^2\left(\frac 12
\ln(\frac{\pi^4T^4K_\star^2}{16})-\frac 12\right)\,. \eqlabel{alg}
\end{equation}
Assuming that $K_\star=K_\star(T)$ and differentiating \eqref{alg},
we find \eqref{leadks}.

\subsubsection{Cascading black hole thermodynamics to order $\calo\left(\frac{P^4}{K_\star^2}\right)$}
We can evaluate the black hole temperature to order
$\calo\left(\frac{P^4}{K_\star^2}\right)$ by requiring that the
Euclidean continuation of the metric \eqref{ktm} does not have a
conical singularity as $y\to 0_+$. We find\footnote{We used from
\eqref{l1sol} $\l_{2,2}^h =\left(-\frac 14+\frac
34\l_{2,0}^h\right)$.}
\begin{equation}
\begin{split}
T=&\frac{2}{\pi}\left(\frac{\ta_0}{K_\star}\right)^{1/2}\biggl\{1-\frac{P^2}{3K_\star}+\frac{P^4}{K_\star^2}\
\left(\frac{(\pi^2+8)^2}{1920}+\frac{1}{30}\left(\l^h_{2,0}\right)^2-\xi^h_{4,0}+2\xi^h_{4,2}-\frac 23 \k^h_{4,0}\right)
\\
&+\calo\left(\frac{P^6}{K_\star^3}\right)\biggr\}\,.
\end{split}
\eqlabel{Tp4}
\end{equation}
Solving for $\ta_0$ from \eqref{Tp4}, and reading off $\{a_0,a_{2,0}\}$ in \eqref{genexp} from \eqref{p2order1}-\eqref{p2order5},
we find
\begin{equation}
\begin{split}
a_0=&\frac 14 T^2\pi^2
K_\star\biggl\{1+\frac{P^2}{2K_\star}+\frac{P^4}{K_\star^2}\
\biggl( \frac{13}{180}+\frac{1}{18}\ln
(2)-\frac{\pi^4}{960}-\frac{\pi^2}{60}-\frac{1}{15}
\left(\l_{2,0}^h\right)^2
+\frac 43 \k_{4,0}^h\\
&+2\xi_{4,0}^h-4\xi_{4,2}^h \biggr)
+\calo\left(\frac{P^6}{K_{\star}^3}\right)\biggr\}\,,
\end{split}
\eqlabel{a04}
\end{equation}
\begin{equation}
a_{2,0}=\frac{7}{48} T^2\pi^2
P^2\left\{1+\frac{P^2}{2K_\star}+\calo\left(\frac{P^4}{K_{\star}^2}\right)\right\}\,.
\eqlabel{a204}
\end{equation}
Additionally we find (see \eqref{num2})
\begin{equation}
P^2 k_s\equiv 4 h_{0,0}a_0^2-\frac 12 P^2 =K_\star\left\{ 1-\frac
{\ln
(2)}{2}\frac{P^2}{K_\star}+\calo\left(\frac{P^6}{K_{\star}^3}\right)\right\}\,.
\eqlabel{K04}
\end{equation}

We are now ready to verify the first law of thermodynamics. The
matching condition \eqref{match1}, to order
$\calo\left(\frac{P^4}{K_{\star}^2}\right)$, gives
\begin{equation}
\calo\left(\frac{P^6}{K_{\star}^2}\right)=4\hh_{0,0}-K_\star+P^2\left(\frac
12 \ln(\frac{\pi^4T^4K_\star^2}{16})-\frac
12\right)+\frac{P^4}{2K_\star}\,, \eqlabel{alg4}
\end{equation}
which results in the following ordinary differential equation for
$K_\star\equiv K_\star(T)$ :
\begin{equation}
\frac{dK_{\star}}{dT}=\frac{2P^2}{T}\left\{1+\frac{P^2}{K_\star}+\calo\left(\frac{P^4}{K_{\star}^2}\right)\right\}\,.
\eqlabel{dk1}
\end{equation}
Now, given \eqref{a04} and \eqref{a204} we can evaluate the energy density $\cale$ and the pressure $\calp$.
The first law of thermodynamics \eqref{first1}
leads to
\begin{equation}
\frac{dK_{\star}}{dT}=\frac{2P^2}{T}\left\{1
+\frac{P^2}{K_\star}\left(\ln (2) -\frac 76 +2\k_{2}^{[4]}\right)
+\calo\left(\frac{P^4}{K_{\star}^2}\right)\right\}\,. \eqlabel{dk2}
\end{equation}
Consistency of \eqref{dk1} and \eqref{dk2} makes a {\it prediction}
\begin{equation}
\k_{2}^{[4]}=\frac{13}{12}-\frac 12\ln(2)\,. \eqlabel{k10p1}
\end{equation}
As a highly non-trivial check on our numerical analysis of the
perturbative expansion, note that \eqref{k10p1} agrees with
\eqref{k20n} to within a factor of order $10^{-10}$.

With \eqref{k10p1} we can also evaluate the speed of sound squared
\begin{equation}
c_s^2\equiv \frac{\del \calp}{\del \cale}=\frac 13 -\frac 49\
\frac{P^2}{K_\star}+\frac{10}{27}\ \frac{P^4}{K_{\star}^2}
+\calo\left(\frac{P^6}{K_{\star}^3}\right)\,. \eqlabel{speedp4}
\end{equation}

\subsubsection{First law of thermodynamics to order $\calo\left(\frac{P^6}{K_\star^3}\right)$}

For the temperature dependence of $K_\star$ to order
$\calo\left(\frac{P^6}{K_\star^3}\right)$ we can again find two
expressions
--- one involving both the UV parameters and the IR parameters, and
the other one parameter-independent. These parallel the expressions
\eqref{dk2} and \eqref{dk1} :
\begin{equation}
\begin{split}
\frac{dK_\star}{dT}=&\frac{2P^2}{T}\ \biggl\{1+\frac{P^2}{K_\star}+\frac{P^4}{K_\star^2}\biggl(
-\frac{1}{480}\pi^4-\frac{1}{30}\pi^2+\frac 19 \ln (2)+\frac{71}{180}-\frac{2}{15}\left(\l^h_{2,0}\right)^2+\frac 83\k^h_{4,0}\\
&+4\xi^h_{4,0}-8\xi^h_{4,2} +\z_2^{[4]}+2\k_2^{[6]}
\biggr)+\calo\left(\frac{P^6}{K_\star^3}\right)\biggr\}\,,
\end{split}
\eqlabel{ksui1}
\end{equation}
\begin{equation}
\begin{split}
\frac{dK_\star}{dT}=\frac{2P^2}{T}\
\left\{1+\frac{P^2}{K_\star}+\frac 12\
\frac{P^4}{K_\star^2}+\calo\left(\frac{P^6}{K_\star^3}\right)\right\}\,,
\end{split}
\eqlabel{ksui2}
\end{equation}
where we used \eqref{k10p}. Consistency of \eqref{ksui1} and
\eqref{ksui2} leads to a prediction
\begin{equation}
-\frac{1}{480}\pi^4-\frac{1}{30}\pi^2+\frac 19 \ln
(2)+\frac{71}{180}-\frac{2}{15}\left(\l^h_{2,0}\right)^2+\frac
83\k^h_{4,0}+4\xi^h_{4,0}-8\xi^h_{4,2}
+\z_2^{[4]}+2\k_2^{[6]}=\frac 12\,. \eqlabel{const}
\end{equation}
We can estimate the error in our solutions by comparing the two
sides of \eqref{const}. Using the explicit expressions for the
perturbative UV parameters \eqref{othern} and the perturbative IR
parameters \eqref{hor1} and \eqref{hor}, we find
\begin{equation}
\bigg|\frac{LHS}{RHS}-1\bigg|\ \sim\ 2\times 10^{-5}\,.
\eqlabel{error3}
\end{equation}

Finally, the speed of sound can be expressed either in terms of the
UV parameters or the IR parameters
\begin{equation}
\begin{split}
c_s^2=&\frac 13 -\frac 49\ \frac{P^2}{K_\star}+\frac{10}{27}\
\frac{P^4}{K_{\star}^2}+\left( -\frac{16}{81}-\frac 49\
\z_2^{[4]}-\frac 89\k_{2}^{[6]} \right)\  \frac{P^6}{K_{\star}^3}
+\calo\left(\frac{P^8}{K_{\star}^4}\right)\,,
\end{split}
\eqlabel{csuv}
\end{equation}
\begin{equation}
\begin{split}
c_s^2=&\frac 13 -\frac 49\ \frac{P^2}{K_\star}+\frac{10}{27}\ \frac{P^4}{K_{\star}^2}+\biggl(
-\frac{1}{1080}\pi^4-\frac{2}{135}\pi^2+\frac{4}{81}\ln (2)-\frac{11}{45}-\frac{8}{135}\left(\l^h_{2,0}\right)^2\\
&+\frac{32}{27}\k^h_{4,0}
+\frac{16}{9}\xi^h_{4,0}-\frac{32}{9}\xi^h_{4,2} \biggr)\
\frac{P^6}{K_{\star}^3} +\calo\left(\frac{P^8}{K_{\star}^4}\right)\,.
\end{split}
\eqlabel{csir}
\end{equation}
Consistency of \eqref{csuv} and \eqref{csir} is guaranteed by \eqref{const}.


\end{document}